\documentclass[10pt,fleqn]{article}

\usepackage{tocloft}

    \usepackage{amsmath,amssymb,amsfonts,amsbsy,amsthm,mathtools}
    \usepackage{latexsym,graphics,graphicx,color,sectsty,mathdots,bm}
    
    \usepackage{subcaption}
    \usepackage{layout}
    \usepackage{bbm}
    \usepackage[colorlinks=true,linkcolor=blue]{hyperref}
    \usepackage{calc,cancel}
    \usepackage{enumitem}
    \usepackage{rotating}
    \usepackage[table]{xcolor}
    \usepackage{cleveref}
    \usepackage[toc]{appendix} 
    \usepackage{sidecap}
    \usepackage{amscd}
    \usepackage{mathtools}
    \usepackage{scalerel}
    \usepackage{relsize}
    \usepackage{mathdots}
    \usepackage{units}
    \usepackage{nicefrac}

    \usepackage{arydshln}
    \usepackage{makeidx} 
    \usepackage{nicematrix}
    \usepackage{fancyhdr}
    \usepackage{xfrac}
    \usepackage{tocvsec2}
    
    \usepackage{mdframed}
    
    \usepackage{upgreek}
    
    \usepackage{tabu}
    
    \usepackage{tcolorbox}

\theoremstyle{plain}
\newtheorem{theorem}{Theorem}
\newtheorem{definition}[theorem]{Definition}
\newtheorem{remarc}[theorem]{Remark}

	\definecolor{bgblue}{rgb}{0.04,0.39,0.53}
	\definecolor{dblue}{rgb}{0,0.3,0.7}
	\definecolor{ddblue}{rgb}{0,0.1,0.6}
	\definecolor{ddgreen}{rgb}{0,0.25,0.05}
	\definecolor{dgreen}{rgb}{0,0.5,0.05}
	\definecolor{mblue}{rgb}{0, .45, .74}
	\definecolor{mgreen}{rgb}{0, .5, 0}
	\definecolor{mred}{rgb}{.7,.2 ,0}

\newcommand{\req}[1]{(\ref{#1.eq})}

\newcommand{\beq}{\begin{equation}}
\newcommand{\eeq}{\end{equation}}
\newcommand{\beqn}{\begin{eqnarray}}
\newcommand{\eeqn}{\end{eqnarray}}
\newcommand{\beqns}{\begin{eqnarray*}}
\newcommand{\eeqns}{\end{eqnarray*}}
\newcommand{\bct}{\begin{center}}
\newcommand{\ect}{\end{center}}
\newcommand{\btmz}{\begin{itemize}}
\newcommand{\etmz}{\end{itemize}}
\newcommand{\benum}{\begin{enumerate}}
\newcommand{\eenum}{\end{enumerate}}

\newcommand{\R}{{\mathbb R}}

\newcommand{\E}{{\mathbb E}}

\newcommand{\sT}{{\scriptstyle T}}

\newcommand{\ut}{{\tilde{u}}}
\newcommand{\xt}{{\tilde{x}}}

\newcommand{\bbm}{\begin{bmatrix}} 
\newcommand{\ebm}{\end{bmatrix}} 
\newcommand{\bmat}{\begin{matrix}} 
\newcommand{\emat}{\end{matrix}}

\newcommand{\bsm}{\left[ \begin{smallmatrix}} 
\newcommand{\esm}{\end{smallmatrix} \right]} 

\newlength{\arraycolseporig} 
\newcommand{\bcrv}{ \arraycolseporig=\arraycolsep \arraycolsep=2pt \begin{bmatrix}} 
\newcommand{\ecrv}{\end{bmatrix} \arraycolsep=\arraycolseporig }

\newcommand{\bsbm}{\left[ \begin{smallbmatrix}} 
\newcommand{\esbm}{\end{smallbmatrix} \right]} 

\newcommand{\bbNm}{\begin{bNiceMatrix}} 				
\newcommand{\ebNm}{\end{bNiceMatrix}} 
\newcommand{\bNA}[1]{ \left[ \begin{NiceArray}{#1} } 		
\newcommand{\eNA}{ \end{NiceArray} \right] }

\newcommand{\lb}{\left(}
\newcommand{\rb}{\right)}
\newcommand{\lcb}{\left\{}
\newcommand{\rcb}{\right\}}

\newcommand{\xinit}{{\sf x_{\rm i}}}

\newcommand{\xinitst}{{\rm x_{\rm i}^*}}

\newcommand{\xfin}{{\rm x_{\rm f}}}
\newcommand{\xfinst}{{\rm x_{\rm f}^*}}

\newcommand{\cK}{{\mathcal K}}

\newcommand{\be}{\begin{equation}}
\newcommand{\ee}{\end{equation}}

\newcommand{\ELtwo}{\sfL^{\!2}}

\newcommand{\cplxs}{ C\kern -.35em \rule{0.03 em}{.7 ex}~   }

\def\complex{\hbox{C\kern -.45em \rule{0.03 em}{1.5 ex}}~}

\newcommand{\xh}{{\hat{x}}}

\newcommand{\wh}{{\hat{w}}}

\newcommand{\rmi}{{\rm i}} 
\newcommand{\rmf}{{\rm f}}

\newcommand{\bi}{\begin{itemize}}
\newcommand{\ei}{\end{itemize}}
\newcommand{\ben}{\begin{enumerate}}
\newcommand{\een}{\end{enumerate}}

\newcommand{\vh}{\hat{v}}

\newcommand{\expct}[1]{  \E  \! \left[ #1 \right]    }

\newcommand{\bseq}{\begin{subequations}}
\newcommand{\eseq}{\end{subequations}}

\newcommand{\ba}{\begin{array}}
\newcommand{\ea}{\end{array}}

\newcommand{\mycaption}[1]{\caption{\footnotesize #1}}
\newcommand{\mysubcaption}[1]{\caption{\scriptsize #1}}

\definecolor{dred}{rgb}{.8,0,0}

\newcommand{\sm}{\text{-}}

\newcommand{\sss}{\scriptscriptstyle}

\newcommand{\ssT}{{\scriptscriptstyle T}}

\def\clap#1{\hbox to 0pt{\hss#1\hss}}

\def\mathrlap{\mathpalette\mathrlapinternal}

\def\mathrlapinternal#1#2{%
           \rlap{$\mathsurround=0pt#1{#2}$}}

\newcommand{\btc}{\begin{tabular}{c}}
\newcommand{\btbl}{\begin{tabular}{l}}
\newcommand{\et}{\end{tabular}}

\newcommand{\fs}{\footnotesize} 
\newcommand{\scz}{\scriptsize}

    \newcommand{\calN}{{\cal{N}}}

    \newcommand{\xd}{\dot{x}}

	\newcommand{\rom}{\rule{0em}{1em}}
	
	\newcommand{\romn}{\rule{0em}{.91em}}

\newcommand{\Nus}[1]{{\sf Nu}\!\left( #1 \right) }

\newcommand{\voTc}{{v_{\scriptscriptstyle [0,T]}}} 
\newcommand{\woTc}{{w_{\scriptscriptstyle [0,T]}}} 
\newcommand{\xoTc}{{x_{\scriptscriptstyle [0,T]}}}

\newcommand{\yoTc}{{y_{\scriptscriptstyle [0,T]}}}

\newcommand{\yotc}{{y_{\scriptscriptstyle [0,t]}}}

\newcommand{\hsom}{\hspace{1em}} 
\newcommand{\hstm}{\hspace{2em}}

\newcommand{\rms}{{\rm s}}

\newcommand{\sfV}{{\sf V}}

\newcommand{\sfW}{{\sf W}}

\newcommand{\sfX}{{\sf X}}

\newenvironment{myenumerate}{\begin{enumerate}[left=0em]}{\end{enumerate}}

\newcommand{\sfO}{{\sf O}}

\newcommand{\sfL}{{\sf L}}

\newcommand{\smint}[2]{{\scaleobj{.8}{\int_{{#1}}^{{#2}}}}}

	\newcommand{\bbms}{\begin{bsmallmatrix}}
	\newcommand{\ebms}{\end{bsmallmatrix}}

               \DeclareMathAlphabet{\mymathbb}{U}{BOONDOX-ds}{m}{n}

	\newcommand{\deffont}[1]{{\sf #1}}

	\newcommand{\mytag}[1]{\tag{\footnotesize #1}}

	\definecolor{lightyellow}{rgb}{1,1,.8}
	\definecolor{lightpurpule}{rgb}{.9,.9,1}

	\newcommand{\RG}{{\,\,\raisebox{.4ex}[0ex][0ex]{\smaller[2]\sf r}\kern-0.49em{\sfO}}}
	\newcommand{\NCG}{{\,\,\raisebox{.4ex}[0ex][0ex]{\smaller[2]\sf n}\kern-0.56em{\sfO}}}
	\newcommand{\OG}{{\,\,\raisebox{.4ex}[0ex][0ex]{\smaller[2]\sf o}\kern-0.54em{\sfO}}}
	\newcommand{\CG}{{\,\,\raisebox{.4ex}[0ex][0ex]{\smaller[2]\sf c}\kern-0.53em{\sfO}}}

	\newcommand{\RGh}{\widehat{\,\,\raisebox{.4ex}[0ex][0ex]{\smaller[2]\sf r}\kern-0.49em{\sfO}}}
	\newcommand{\OGh}{\widehat{\,\,\raisebox{.4ex}[0ex][0ex]{\smaller[2]\sf o}\kern-0.54em{\sfO}}}

	\newcommand{\speq}{\, = \, }

	\newcommand{\smfm}[1]{\mbox{\fs $#1$}}

	\newcolumntype{M}[1]{>{\centering\arraybackslash}m{#1}}

	\newenvironment{mytcolorbox}{\medskip\begin{tcolorbox}[colframe=gray,boxrule=.1pt,left=1mm,right=1mm,boxsep=0mm,before skip=2pt,after skip=6pt]}{\end{tcolorbox}}
	\newenvironment{mytcolorboxtitle}[1]{\medskip\begin{tcolorbox}[colframe=gray,boxrule=.1pt,title={\sf #1},before skip=2pt,after skip=6pt,left=1mm,right=1mm]}{\end{tcolorbox}}

	\newcommand{\bbmsh}{\renewcommand{\arraystretch}{1} \begin{matrix}} 		
	\newcommand{\ebmsh}{\end{matrix}}

	\newcommand{\highcolbox}[1]{\colorbox{lightgray!30}{$#1$}}
	\newcommand{\emphbox}[1]{\colorbox{lightgray!30}{#1}}

	\MakeAboxedCommand\Ahighcolbox\emphbox

	\newcommand{\Qba}{\bar{Q}}
	\newcommand{\Aba}{\bar{A}}
	\newcommand{\Bba}{\bar{B}}
	\newcommand{\Pba}{\bar{P}}
	\newcommand{\Sba}{\bar{S}}
	
	\newcommand{\sfXba}{\bar{\sfX}}
	\newcommand{\tbd}{t_{\rm b}}

\newcommand{\LQRi}{{\sf LQR_i}}
\newcommand{\LQRf}{{\sf LQR_f}}

\newcommand{\bsmtwo}[4]{\mbox{\tiny $\arraycolsep=2pt
				\left[ \begin{array}{ccc:c}  & & &					 \\
					 ~& \mbox{\normalsize $#1$} &  ~& \mbox{\normalsize $#2$}  		\\ 
					 	& & & 							\\
						 \hdashline &  \mbox{\normalsize $#3$}  & & \mbox{\normalsize\romn $#4$}  
				\end{array}  \right] $}}
\newcommand{\bsmtbo}[2]{\mbox{\tiny $ \arraycolsep=2pt
	\left[ \begin{array}{c} ~ \\ \mbox{\normalsize $#1$} \\ ~ \\ \hdashline \mbox{\normalsize \rom$#2$} 
		\end{array} \right] $}}
\newcommand{\bsmobt}[2]{\mbox{\tiny $ \arraycolsep=2pt
	\left[ \begin{array}{ccc:c} &  \mbox{\normalsize $#1$} & & \mbox{\normalsize \rom$#2$} 
		\end{array} \right] $}}
\newcommand{\shorteq}{\mathrel{\vcenter{\hbox{\scalebox{0.75}[1]{$=$}}}}}

\title{Least-Squares State Estimation, LQR and LQ-Tracking}

\author{Bassam Bamieh\thanks{Department of Mechanical Engineering, 
	and Center for Control, Dynamical Systems, and Computation (CCDC), 
	University of California at Santa Barbara, {\em bamieh@ucsb.edu.} }}

\date{}

\begin{document}

\maketitle

\begin{abstract} 
		This note is a tutorial on the Least-Squares State Estimator (LSSE) (the deterministic version of the  Kalman-Bucy filter) and related topics. The LSSE is  formulated as finding the state trajectory consistent with the system's equations with the minimal amount of $\ELtwo$ process and measurement uncertainty. As stated, this is an input-signal design problem with linear dynamics and affine-quadratic objective in the state and inputs, and  therefore a deterministic optimal control problem. We explore its relations to other problems such as the Linear Quadratic Regulator (LQR) with initial or final conditions, as well as the Linear Quadratic (LQ)-tracking problem. Several related topics such as the use of homogeneous coordinates and time reversal in optimal control are explored. The emergence of dynamical controllers/estimators in both LQ-tracking and LSSE  as opposed to memoryless ones (as in LQR) is highlighted. It is seen to be a consequence of the affine-quadratic,  rather than a purely quadratic form of the cost objective.  The relations with the stochastic version of the Kalman-Bucy filter are explicitly highlighted, as well as characterizations in terms of certainty (information) matrices, versus covariance matrices. 
\end{abstract} 

%
%
%

\section{Introduction and Problem Formulation}

	The Kalman-Bucy filter was originally  formulated~\cite{kalman1960new,kalman1961new} 
	as a stochastic state estimation problem 
	in the presence of uncertainty in the dynamics, measurements and initial states, all of which 
	are characterized probabilistically using their second-order statistics. 
	On the other hand, it
	has long been known that minimum-variance type estimation problems also have 
	equivalent deterministic least-squares
	 versions. Which version (deterministic versus stochastic) of these problems 
	one prefers is often a matter of taste rather than logical necessity. Indeed, as 
	Willems~\cite{willems2002deterministic} argues ``This has been 
	a matter of debate at least since Gauss justified Legendre's least squares as a method of 
	computing the most probable, maximum likelihood, outcome". 
	

		In this tutorial note, a motivation for the deterministic Kalman filter is given 
	in line with 
	Willems' arguments~\cite{Willems2004DeterministicLeastSquaresFiltering,willems2002deterministic}.
	The derivation given here follows more closely that of 
	Sontag~\cite[Sec. 8.3]{sontag2013mathematical}, who uses the equivalence with the 
	LQ-tracking problem, combined with a time reversal and a further optimization 
	of the value function over the final state. This approach 
	 has the advantage of showing more clearly the connections
	between LQR, LQ-tracking and the present least-squares state estimation problem. 
	A more streamlined version of this 
	argument is presented  which disentangles the several steps, namely time reversal, 
	introduction of homogeneous coordinates, and the further optimization of the 
	value function.

		\subsection*{Problem Formulation}

	We now state the  motivation  and formulation of  the deterministic 
	version of the state estimation problem as advocated by 
	Willems~\cite{Willems2004DeterministicLeastSquaresFiltering}
	(see also Hespanha's textbook~\cite[Sec. 24.2]{hespanha2018linear}, as well 
	as~\cite{aguiar2003minimum,aguiar2006minimum}). 
	For simplicity we consider a system model without a control input, since the case
	with a control input is essentially identical. The uncertain system model is 
	\be
		\begin{aligned} \xd(t) &\speq A \, x(t) + w(t) , \\ 
							y(t) &= C \, x(t) + v(t) , 
		\end{aligned} 
		\hstm x(0) = \xinit ,  \hstm t\in[0,\sT].
	  \label{dyn_unc.eq}
	\ee
	We use the notation $w_{\sss [0,T]} := \lcb w(t); ~ t\in[0,\sT] \rcb$ to refer to the entire 
	signal over the time horizon $[0,\sT]$ when that is needed for emphasis. 
	We call the system~\req{dyn_unc} ``uncertain'' because the only signal that is known 
	is the output $\yoTc$. All the other signals $\woTc$, $\voTc$, $\xoTc$, and the
	initial state vector $\xinit\in\R^n$ are unknown. Given an observation 
	$\yoTc$, there is an infinite number of combinations of the unknown quantities 
	that satisfy the dynamics~\req{dyn_unc}. How should one choose {\em the best} 
	from among those? 

	To guide this choice, first note that as mentioned, the uncertainty in the 
	system~\req{dyn_unc} is ``parametrized'' by $(\woTc,\voTc,\xoTc,\xinit)$. However, 
	this is an overparameterization 
	since $\xinit$ and $\woTc$ together determine $\xoTc$. Therefore,  we can choose 
	to parameterize the uncertainty with just the triple $(\woTc,\voTc,\xinit)$. 
	Whatever parameterization of the uncertainty we use, the key idea is as follows. 
%
	\begin{mytcolorboxtitle}{Minimal Uncertainty Principle}  
		Given the observation $\yoTc$, the optimal estimate $\xh_{\sss [0,T]}$ of the state trajectory 
			 is the signal consistent with the system equations~\req{dyn_unc}, 
			with the minimal ``size'' of the uncertainty triple $(\woTc,\voTc,\xinit)$.
	\end{mytcolorboxtitle} 
	\noindent
	Note that this principle leaves the notion of uncertainty ``size'' as a choice. 
	The most 
	mathematically tractable choice made below uses  quadratic norms. 
	
	The 
	principle stated above can be thought of as a version of ``Occam's razor'', which roughly 
	speaking is: {\em among all the possible explanations, choose the one
	with the fewest assumptions}. Here the ``assumptions'' are the uncertain (unknown)  
	triple $(\woTc,\voTc,\xinit)$, and choosing 
	the smallest size uncertainty is a proxy for ``fewest assumptions''. 
	
	As already mentioned, the most mathematically tractable choice of uncertainty 
	size is a quadratic functional of the form 
	\be
		J(v,w,\xinit) := \smint{0}{T} \lb v^*\sfV\, v + w^* \sfW \, w \romn \rb dt + \xinitst \sfX \, \xinit
			=: \| \sfV^{\frac{1}{2}} v \|_{\ELtwo}^2 + \| \sfW^{\frac{1}{2}} w \|_{\ELtwo}^2 
					+ \|	\sfX^{\frac{1}{2}}  \xinit	\|_{\R^n}^2 , 
	  \label{MSE_cost.eq}
	\ee
	where $\| \sfV^{\frac{1}{2}} v \|_{\ELtwo}$, $\| \sfW^{\frac{1}{2}} w \|_{\ELtwo}$ are weighted $\ELtwo[0,\sT]$ 
	norms of the signals $v$, $w$ respectively (weighted by the matrices $\sfV$ and $\sfW$ respectively), 
	and $\| \sfX^{\frac{1}{2}} \xinit\|_{\R^n}^2$ is a weighted Euclidean 
	norm of the vector $\xinit$. The weight matrices $\sfV, \sfW, \sfX$ are design choices made
	based on the designer's a priori assumptions about the relative sizes of the various 
	uncertainty components\footnote{The initial-state cost can also be generalized 
		to $\|\xinit - \bar{x}\|_\sfX^2$ where $\bar{x}$ is a prior mean. We omit 
		this for simplicity of notation.}. 
	The standing assumption here is that the uncertain signals 
	$v,w$ belong to $\ELtwo[0,\sT]$. 
	
	The relations between the weight matrices $\sfV,\sfW,\sfX$ and the more familiar 
	noise and initial state covariance matrices are described in Section~\ref{KF_cov.sec}, where 
	the connections between deterministic and stochastic versions of the Kalman filter 
	are discussed. 
	
	Before stating the problem formally, observe that the output equation~\req{dyn_unc} can be 
	used to express the $v$ component of the cost $J$ in terms of $x$ and $y$ as follows 
	\[
		\lb y = C \, x + v 
		\hsom \Leftrightarrow \hsom 
		v = y - Cx  \romn	\rb 
		\hstm \Rightarrow \hstm 
		\smint{0}{T} (v^* \sfV \, v ) \, dt = \smint{0}{T} (y-Cx)^{\!*}\,  \sfV \, (y-Cx) \, dt . 
	\]
	Now the problem can be stated formally.
	\begin{mytcolorboxtitle}{($\sfL^2$)  Least-Squares State Estimation (LSSE)} 
		Given the measured signal $\yoTc$, find
		 the optimal signals $\xoTc$, $\woTc$ 
		satisfying  
		\[
			\dot{x}(t) \speq A \, x(t) +  w(t) 
		\]
		that minimize the cost 
		\be
			J(x,w) := \smint{0}{T} \big(   (y-Cx)^{\!*}\,  \sfV \, (y-Cx)  + w^* \sfW \, w \big) \, dt 
				+ x^*\!(0) \, \sfX \, x(0) . 
		  \label{l_uncert.eq}
		\ee
	\end{mytcolorboxtitle} 
	\noindent 
	As stated, this is an optimal ``input design'' problem. The solution gives not only the optimal 
	state estimate $\xh$, but also the optimal inputs $\wh$ and $\vh$ (recovered from 
	$\vh = y - C \xh$). It is closely related to
	the LQ-tracking problem, though there is an important difference in that the initial state is 
	also unknown. It is also different from an LQR problem in that the cost $J$ is not 
	purely quadratic in the state and input, but rather quadratic+linear+constant
	(also called ``affine-quadratic'') in the state since 
	\be
		(y-Cx)^{\!*}\,  \sfV \, (y-Cx) 
		\speq 
			\underbrace{x^*C^*\sfV C \,  x }_{\mbox{\fs quadratic in $x$}}
			- \underbrace{y^*\sfV C \, x - x^* C^*\sfV y}_{\mbox{\fs linear in $x$}}
			+ \underbrace{y^* \sfV \, y}_{\mathrlap{\mbox{\fs constant in $x$}}}. 
	  \label{aff_quad_first.eq}
	\ee

	\subsection*{Outline of the Paper} 
	

		The starting point is the traditional Linear Quadratic Regulator (LQR) for a possibly 
	time-varying system over a finite horizon. It is assumed that the reader is familiar 
	with this problem and its solution via a matrix Differential Riccati Equation (DRE). 
	A ``dual'' problem (unrelated to estimation) is also formulated where the final, 
	rather than initial, state is specified. This leads to a DRE that characterizes the 
	``cost-to-arrive'' function rather than the cost-to-go. This dual problem is of interest 
	in its own right and forms the basis for the optimal estimator equations developed later. 
	However, the dual problem as stated is a control problem 
	(or more precisely, an input-design problem) rather than an estimation problem.  
	We call this problem ``LQR with final conditions''. Its solution can be easily derived 
	from the traditional  LQR with initial conditions problem via a time reversal as 
	presented in Section~\ref{LQR.sec}. 
	
	The  state-estimation problem as formulated above is an 
	affine-quadratic
	optimal control problem. It can be embedded   in  a larger  {\em purely quadratic} problem  
	(i.e. of the LQR type)
	by appending  an additional scalar state that is enforced to always 
	equal  $1$. This technique is akin to the introduction of so-called 
	``homogeneous coordinates'' common in optimization, computer graphics and 
	projective geometry. It is of interest in its own right and is described carefully 
	in Section~\ref{homo.sec}. 
	
	The next step in Section~\ref{homo.sec} 
	 is to partition the Differential Riccati Equation (DRE) of the larger purely 
	quadratic LQR problem, and extract three differential equations from it. 
	One of those is a DRE for the original problem, and the 
	second is a linear system of the same state dimension as the original system. 
	This reveals why 
	controllers for affine-quadratic cost problems contain a dynamical system, 
	while controllers for purely quadratic costs are memoryless. In the LQ-tracking 
	problem (Section~\ref{tracking.sec}), 
	those dynamics are  the anti-causal feedforward part of the control, while 
	in the estimation problem (Section~\ref{st_est.sec}), the dynamics produce 
	 those of the causal observer! 
	
	There is an additional  step which appears only in the estimation problem. 
	In Section~\ref{st_est.sec}, 
	the estimation problem is solved as if the final state is known. 
	The final state however is unknown, but this solution falls out of the
	LQR problem with final conditions rather easily given the so-called 
	cost-to-arrive function. It is then very simple to further optimize the cost-to-arrive 
	with respect to the (unknown) final state to yield the Kalman filter equations. 

	

	In the stochastic version of the state estimation problem, the DRE solution represents the optimal 
	error covariance matrix. The DRE derived here in the deterministic case turns out to be for 
	the inverse of the covariance matrix. Section~\ref{KF_cov.sec} gives geometric  interpretations and 
	comparisons of those matrices. In particular, the deterministic DRE is interpreted as providing 
	a ``certainty matrix'' which describes how ``certain'' the optimal estimator is of its estimate in 
	comparison to other candidates.  Geometrically, those matrices give ellipsoids that describe 
	relative certainty and uncertainty of different directions in state space. In particular, the 
	certainty matrix is characterized without probabilistic language, as the 
	{\em curvature of the cost-to-arrive
	about its minimizer}. 
	
	Finally, some comments about
	the larger context of developments in this paper are given in the Discussion Section~\ref{disc.sec}.

\section{The LQR Problem with Initial or Final Conditions} 					\label{LQR.sec}

	We consider two versions of the  standard LQR problem over a finite time horizon $[0,\sT]$. One version 
	is where the initial condition is specified. This is the most commonly used version. However, there is a 
	parallel version where the final condition is specified instead. Specifically, we consider the 
	following two problems. 
	
	\begin{myenumerate} 
	\item {\sf LQR with initial conditions:} This is the standard LQR problem over a finite time horizon with 
		 a specified initial condition and a final state penalty
                    	\be
				{\sf LQR_i}: \hstm 
                    			\begin{aligned} 	\xd &= A \, x + B \, u, 	\\ 
                    								 x(0) &= \xinit, 
					\end{aligned} 
                    		\hstm  
                    		J(u,\xinit) =  \smint{0}{T}\!\! \lb x^*Q\, x+ u^* R \, u \rb dt + \|  x(\sT) \|^2_\sfX , 
                    	  \label{LQR_init.eq}
                    	\ee
			where $\|x(\sT)\|^2_\sfX := x^{\!*}\!(\sT) \, \sfX \, x(\sT)$ is the final state penalty.  
			As is well known~\cite{kirk1970optimal,sage1977optimum}, 
                    	the optimal input is obtained from the solution of a matrix Differential Riccati Equation (DRE) 
                    	with final boundary conditions
                    	\be
                    		u \speq -R^{\sm1} B^*P \, x, \hstm \hstm 
                    		-\dot{P} \speq A^*P + PA - PBR^{\sm1} B^* P + Q, 
                    		\hstm P(\sT) = \sfX, 
                    	  \label{DRE_init.eq}
                    	\ee
                    	with the quadratic value function 
                    	\[
                    		\inf_u \, J(u,\xinit)  \speq \xinitst P(0) \, \xinit .
                    	\]
			This is referred to as the ``cost-to-go'' evaluated at $t=0$. 
	\item {\sf LQR with final conditions:} 
                	On the other hand, an LQR problem with a final state target $\xfin$ and an initial state penalty
		 can also be formulated  as follows
                	\be
			{\sf LQR_f}: \hstm
                		\begin{aligned} 	\xd &= A \, x + B \, u, 	\\
                								 x(\sT) &= \xfin,
			\end{aligned}  
                		\hstm  
                		J(u,\xfin) =  \smint{0}{T} \!\! \lb x^*Q\, x+ u^* R \, u \rb dt  + \| x(0) \|^2_\sfX , 
                	  \label{LQR_fin.eq}
                	\ee
		where $\|x(0)\|^2_\sfX := x^{\!*}\!(0) \, \sfX \, x(0)$ is the initial-state penalty. 
               The optimal solution to this problem  is also obtained from a DRE, but one with 
                {\em 	initial boundary conditions}
                	\be
                		u \speq R^{\sm1} B^*S \, x, \hstm \hstm 
                		\dot{S} \speq -A^*S - SA - SBR^{\sm1} B^* S + Q, 
                		\hstm S(0) = \sfX, 
                   \label{DRE_fin.eq}
		\ee
                	with the optimal value as  
                	\[
                		\inf_u \, J(u,\xfin)  \speq \xfinst S(\sT) \, \xfin .
                	\]
		This is sometimes referred to as the ``cost-to-arrive'' in the Moving Horizon Estimation (MHE)
		literature~\cite{diehl2014lecture}.
	\end{myenumerate} 
	
	Note that the main difference between the solutions of the two problems 
	 is the final~\req{DRE_init} versus initial~\req{DRE_fin} 
	condition on the DRE. 
	The solution~\req{DRE_fin} can be derived  from the standard minimum principle arguments that lead to 
	a Two Point Boundary Value Problem (TPBVP) with a linear relation between the state and co-state at initial time. 
	Propagating that relation forward in time gives the DRE~\req{DRE_fin}. Alternatively, it can be derived 
	from the $\LQRi$ problem by a simple time reversal as follows. Define the time-reversed 
	signals 
	\[
		\xt (t) := x(\sT-t), \hstm \ut(t) := u(\sT-t). 
	\]
	The problem~\req{LQR_fin} then becomes 
                	\be
			{\sf LQR_f}: \hstm
                		\begin{aligned} 	\dot{\xt} &= -A \, \xt - B \, \ut, 	\\
                								 \xt(0) &= \xfin,
			\end{aligned}  
                		\hstm  
                		J(\ut,\xfin) =  \smint{0}{T} \!\! \lb \xt^*Q\xt+ \ut^* R \ut \rb dt  + \| \xt(\sT) \|^2_\sfX.
                	  \label{LQR_fin_revs.eq}
                	\ee
	Note that the integral part of the cost is unchanged. Now apply the $\LQRi$ solution~\req{DRE_init} 
	to this problem, and define the matrix function $S(t):= P(\sT-t)$ as the time-reversed solution of the 
	DRE~\req{DRE_init}. This gives the solution~\req{DRE_fin} after another time reversal to obtain the 
	original $x$ and $u$ 
	signals. 
	
	\begin{remarc} 						\label{Q_R_X.remark}
		Since these are finite-horizon problems, sufficient conditions  for the existence of 
		solutions to the DREs~\req{DRE_init} and~\req{DRE_fin} over $[0,\sT]$ are $Q, \sfX \geq 0$ and $R>0$. 
		For positive definiteness of the solution, several additional conditions can guarantee this. 
		For example $\sfX>0$ guarantees $P(t)>0$ for all $t\in[0,\sT]$. Also, $(Q,A)$ observable guarantees
		$P(t)>0$ in $[0,\sT]$ except possibly at boundary points ($t=0$ or $t=\sT$).
		These statements are usually derived through analyzing the 
		properties of the DREs (the so-called ``Riccati comparison'' results), and other system-theoretic 
		results. 
		\end{remarc} 
	
		In the LQ-tracking and state estimation problems discussed below, 
		the conditions above 
		translate to other conditions on the problem data in each case as will be shown.
		Note that stabilizability of $(A,B)$ and detectability of $(Q,A)$ are only needed for infinite-horizon 
		problems since stability is an asymptotic concept. 
	
%

	\section{Homogeneous Coordinates in Optimal Control} 				\label{homo.sec}
	
	\begin{definition} 
	A functional $f:\R^n\rightarrow \R$ which is the sum of  quadratic, linear and constant terms 
	has the general form 
	\be
		f(x) \speq x^*Q \, x + q_1^* x + x^* q_1 + q_0 , 
		\hstm \hstm 
		Q=Q^*, ~	q_1\in\R^n, ~q_0\in\R. 				\label{f_x_def.eq}
	\ee
	We call such a  functional \deffont{affine-quadratic}. 
	\end{definition} 
	\noindent 
	Note that the two linear terms are equal $q_1^*x + x^* q_1 = 2\, q_1^* x = 2 \, x^* q_1$, but we write them 
	as above for the sake of symmetry. 
	
\begin{figure}[t]
		\centering
		\includegraphics[width=0.9\textwidth]{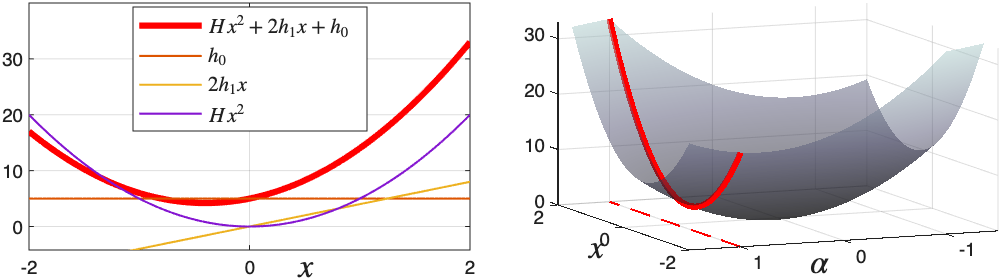} 
		
		\mycaption{Conversion of an affine-quadratic functional into a purely quadratic functional 
			in homogeneous coordinates. ({\em Left}) An affine-quadratic functional is the sum of a 
			quadratic, linear, and constant functionals in $x$. ({\em Right}) By appending a scalar 
			variable $\alpha$, the affine-quadratic functional becomes the value of a purely quadratic 
			functional~\req{homo_vect}  
			in the variables $(x,\alpha)$  over the set $\alpha =1$. 
			} 
	  \label{Homogenize.fig}
	\end{figure}

	By introducing an auxiliary variable equal to $1$, an affine-quadratic functional  can be converted into a purely 
	quadratic functional by defining a vector with $n+1$ components as  
	\be
		\xi := \bsmtbo{x}{1}  , ~
		\Qba := \bsmtwo{Q}{q_1}{q_1^*}{q_0} , 
		\hstm \hstm 
		f(x) \speq 	
			\bsmobt{x^*}{1} 
			\bsmtwo{Q}{q_1}{q_1^*}{q_0} 
			\bsmtbo{x}{1}
			\speq \xi^* \Qba \, \xi .
	  \label{homo_vect.eq}
	\ee
	This is a common construction in optimization, computer graphics, and projective geometry. 
	See Figure~\ref{Homogenize.fig} for an illustration. 
	\begin{remarc} 
		In the case $Q>0$, this is related to the method of ``completing the square''. Indeed, examining 
		Figure~\ref{Homogenize.fig} we see that the affine-quadratic functional looks like a quadratic 
		functional (with the same curvature as the original), but centered at a different abscissa and 
		shifted vertically. The minimizer  of $f$ in~\req{f_x_def} is $\xh = -Q^{\sm1} q_1$ with 
		minimum value $f(\xh) = q_0 -q_1^* Q^{\sm1} q_1 $. Using this,  $f$ can be rewritten as 
		\begin{align*}  
			(x-\xh)^* Q \, (x-\xh) + ( q_0 -  q_1^* Q^{\sm1} q_1) 
			&\speq 
			(x+Q^{\sm1} q_1)^* Q \, (x+ Q^{\sm1} q_1 ) + ( q_0 -  q_1^* Q^{\sm1} q_1) 		\\
			&\speq 
			x^* Q \, x  + q_1^* x + x^* q_1+  q_0 = f(x). 
		\end{align*} 
		Thus $f(.)$ can be rewritten as a ``perfect square'' $(x-\xh)^* Q \, (x-\xh)$ plus a constant. 
		However, the construction~\req{homo_vect} is more general since it does not require $Q>0$, 
		which is often not the case in optimal control problems.  
	\end{remarc} 
	
	
	The construction~\req{homo_vect} can also be used for optimal control problems where the objective is
	affine-quadratic. The idea is to enlarge 
	the state space by one dimension, and  append a scalar  state that 
	is always equal to $1$. This constant state 
	can then be  used  as the  term equal to $1$ in the homogeneous coordinates. For example, 
	consider the following linear  dynamics and affine-quadratic objective
	(temporarily omitting initial/final state constraints and penalties) 
	\be
            	\begin{aligned} 
            		\xd =& \, A \, x + B \, u , 
            						\\
            		J =& \smint{0}{T} \lb x^* Q \, x + q_1^* x + x^* q_1 + q_0 \, +\,  u^* R \, u  \rom \rb dt   
				= \smint{0}{T} \!\!  \lb 
					\smfm{\arraycolsep=3pt  
					  \bbm x \\ 1 \ebm^{\!*}\! \bbm Q & q_1 \\ q_1^* & q_0 \ebm \bbm x \\ 1\ebm}
					+ u^*R \, u \rb dt 
            	\end{aligned} 
	  \label{qla_prob.eq}
	\ee
	where the matrices $A,B,Q,R$,  the vector $q_1$, and the scalar $q_0$ may be functions of time. 
	The  problem~\req{qla_prob} is equivalent to  the following problem with a larger state space 
	\be
		\begin{aligned}
            		\bsmtbo{\xd}{\dot{\alpha}} 
            		&= 
            		\bsmtwo{A}{0}{0}{0} \bsmtbo{x}{\alpha} + \bsmtbo{B}{0} \, u , 	\\ 
            		\alpha (0) & =1,  ~\mbox{or} ~ \alpha(\sT) = 1, 
		\end{aligned} 
		 \hstm  \Leftrightarrow \hstm 
		 \dot{\xi} \speq \Aba \, \xi + \Bba \, u , 
		 \hstm 
		 \xi :=  \bsmtbo{x}{\alpha},
	  \label{A_B_part.eq}
	\ee
	The state $\alpha$ is then equal to $1$ for all $t$, and therefore  can act like the constant $1$ in the homogeneous 
	state coordinates $\xi$. 
	The objective~\req{qla_prob} now   becomes 
	\begin{align*} 
		J = \int_{0}^{T} \!\! \lb \arraycolsep=3pt   {  \bbm {x} \\ {\alpha}\ebm^* 
					\bbm {Q} & {q_1} \\ {q_1^*} & {q_0} \ebm  
					\bbm {x} \\ {\alpha} \ebm 	
					+ u^* R \, u } \rule{0em}{1.5em} \rb  dt  
			= \smint{0}{T} \!\! \lb \xi^* \Qba \, \xi + u^* R \, u \rb dt , 
	\end{align*} 
	which is a  purely quadratic cost. This is stated next together with initial/final state constraints 
	and penalties. 
%
	\begin{mytcolorboxtitle}{Equivalence of Affine-Quadratic Problem~\req{aff_quad} and (Purely) Quadratic Problem~\req{quad_embed}} 	\vspace{-1em}
	\be 
		\hspace{-.5em} 
			\xd =  A \, x + B \, u , 
			~ 
		J \!=\! \smint{0}{T} \!\!  \lb 
					\smfm{\arraycolsep=3pt  
					  \bbm x \\ 1 \ebm^{\!*}\! \bbm Q & q_1 \\ q_1^* & q_0 \ebm \bbm x \\ 1\ebm}
					  + u^*\!R u \rb dt    
			+ \| x(\tbd) \|_\sfX^2 , 
		~ 
		\lb \begin{aligned} x(0) & \! \shorteq \!  \xinit, \,  \tbd \! \shorteq \! \sT, \\ 
				\!\mbox{\raisebox{.8em}[0em]{or}}
					 x(\sT) & \! \shorteq \! \xfin, \,  \tbd \! \shorteq \! 0 . 
			\end{aligned} \rb 
	  \label{aff_quad.eq}											
	\ee
	\vspace{-.5em}
	\be
            	\begin{aligned} 
            		\dot{\xi} =& \Aba \, \xi + \Bba \, u ,  
				&  \Aba \!:\shorteq& \smfm{\bbm A & 0 \\ 0 & 0 \ebm}, 
					\, \Bba:= \smfm{\bbm B \\ 0 \ebm} , 				\\
         		J =& \smint{0}{T} \!\! \big( \xi^* \!\Qba  \xi + u^*\! R  u \big) dt + \| \xi(\tbd) \|_{\sfXba}^2 , 
				& \hspace{-1em} \Qba \! :\shorteq&\arraycolsep=2pt  \smfm{\bbm Q & q_1 \\ q_1^* & q_0 \ebm, \,  
										\sfXba := \! \bbm \sfX & 0 \\ 0 & 0 \ebm,} 
            	\end{aligned} 
		\lb  
		\begin{aligned} 
				\xi(0) \!&\shorteq\!  	\smfm{ \bbm {\xinit} \\ {1}  \ebm} , \,  \tbd\! \shorteq \!\sT, \\ 
            			\!\mbox{\raisebox{1.2em}[0em]{or}}\!\! \xi(\sT) \!&\shorteq\! \smfm{ \bbm {\xfin} \\ {1}\ebm }   , \, \tbd\! \shorteq\! 0. 
		\end{aligned}			\rb
	 \label{quad_embed.eq}
	\ee
	\end{mytcolorboxtitle}
	\noindent
	Note the two versions of the problem where if the ``boundary penalty'' is at  $\tbd = \sT$, 
	then an initial  state is specified $x(0)=\xinit$, 
	 and vice versa if the boundary penalty is at  $\tbd=0$, then a final state $x(\sT)=\xfin$ is specified.

	Provided existence conditions are satisfied, the 
	 purely quadratic  problem~\req{quad_embed} is solvable as a standard 
	(possibly time-varying) LQR problem with either initial or final conditions as 
	described in Section~\ref{LQR.sec}. 
	The partitioning above of the $\Aba, \Bba, \Qba$ matrices induces 
	a partitioning of the DRE solution as well as the optimal input $u$ as  shown next. 
	
%
	
	\subsection{Affine-Quadratic Problems with Initial Conditions} 

	In the case of specified initial conditions, the
	 DRE for the  homogenized problem~\req{quad_embed} with initial 
	 conditions is given by~\req{DRE_init}. Assuming  existence conditions hold (see Remark~\ref{H_R_init.remark}
	 below), label  
	its solution as $\Pba$ and partition it conformably with the partitions in~\req{A_B_part} as follows 
	\be
		-\dot{\Pba} = \Aba^*\Pba + \Pba \Aba - \Pba \Bba R^{-1} \Bba^* \Pba + \Qba , 
		\hstm \hstm \Pba = \smfm{ \bbm P & p_1 \\ p_1^* & p_0 \ebm }, 
		\hstm \Pba(\sT)  = \smfm{ \bbm \sfX & 0 \\ 0 & 0 \ebm }. 
	  \label{Ph_eq.eq}
	\ee
	Note that $P$ is an $n\times n$ matrix (the state dimension of the original problem), while 
	the off-diagonal term $p_1$ is an $n$-vector, and $p_0$ is a scalar. 
	The partitioning of the $\Aba, \Bba, \Qba$ matrices gives the following 
	set of equations
		\begin{align*} 
			\hspace{-2em}
			{\arraycolsep=3pt
			\sm \smfm{ \bbm \dot{P} & \dot{p}_1 \\ \dot{p}_1^* & \dot{p}_0 \ebm } }
			=\,& {\arraycolsep=3pt
				\smfm{\bbm A^* & 0 \\ 0 & 0 \ebm } 
				 \smfm{ \bbm P & p_1 \\ p_1^* & p_0 \ebm } + 
				 \smfm{ \bbm P & p_1 \\ p_1^* & p_0 \ebm }
				  \smfm{\bbm A & 0 \\ 0 & 0 \ebm } 
				  -
				  \smfm{ \bbm P & p_1 \\ p_1^* & p_0 \ebm } 
				  \smfm{ \bbm B \\ 0 \ebm } R^{\sm 1} \smfm{ \bbm B^* & 0 \ebm } 
				  \smfm{ \bbm P & p_1 \\ p_1^* & p_0 \ebm } 
				  + 
				  \smfm{\bbm Q &  q_1 \\ q_1^* & q_0 \ebm} 	}		\\
			=\,& 
				{\arraycolsep=3pt
				\smfm{\bbm A^*P & A^*p_1 \\ 0 & 0 \ebm } 
				 + 
				  \smfm{\bbm PA & 0 \\ p_1^*A  & 0 \ebm } 
				  -
				  \smfm{ \bbm PBR^{\sm1}B^* P & PBR^{\sm1} B^* p_1  \\ 
				  			p_1^*BR^{\sm1}B^* P & p_1^*BR^{\sm1}B^* p_1 \ebm    } 
				  + 
				  \smfm{\bbm Q &  q_1 \\ q_1^* & q_0 \ebm} 	}.
		\end{align*} 
		This gives  three equations for the (1,1), (1,2), and (2,2) 
		blocks respectively
		\begin{mytcolorbox}		\vspace{-1.3em}
		\begin{align} 
			- \dot{P} \speq& A^* P + P A - P BR^{-1} B^* P + Q, 
				& P(\sT) &= \sfX,  							\label{P1_DRE.eq}	\\ 
			- \dot{p}_1 \speq& A^* p_1 - P BR^{-1} B^* p_1 + q_1 , 
														\nonumber 		\\
					  \speq& (A + BK )^*\,   p_1 + q_1 , 
					  \hstm K := -R^{-1}B^* P, 
				& p_1(\sT) & = 0, 							\label{po_DRE.eq}	\\
			- \dot{p}_0 \speq& - p_1^*BR^{-1}B^* p_1 + q_0 , 
				& p_0(\sT) &= 0, 							\label{p2_DRE.eq}
		\end{align} 
		\end{mytcolorbox}
		\noindent 
		Together, these three equations are equivalent to~\req{Ph_eq}.

		Equation~\req{P1_DRE} is a DRE associated with a standard LQR problem with the matrices $(A,B,Q,R)$. 
		Equation~\req{po_DRE} is a linear system for the vector signal $p_1$, driven by the 
		constant or time-varying $n$-vector  $q_1$, the off-diagonal block of the matrix $\Qba$. 
		Note that the ``$A$-matrix'' of this linear system depends on the solution $P$ of the DRE~\req{P1_DRE}.
		We will see that in the LQ-tracking problem,  
		the linear system~\req{po_DRE} for $p_1$  gives the (anti-causal) feedforward part of the control. 
				
		The optimal input $u$ is given by the standard formula~\req{DRE_init} 
		 applied to  the  problem~\req{quad_embed} 
		 \begin{align} 
		 	u = -R^{\sm1} \Bba^* \Pba \, \xi 
				= -R^{\sm1} \bbm B^* & 0 \ebm 
					\smfm{ \bbm P & p_1 \\ p_1^* & p_0 \ebm } 
					\smfm{\bbm x \\ 1 \ebm} 
				&= -R^{\sm1} B^* \lb P \, x + p_1 \rb 			\nonumber			\\
				 & = {-R^{\sm1} B^*  P \, x} 
							-  {R^{\sm1} B^* p_1}   .
															 \label{u_opt.eq}	
		 \end{align} 	
		 Note that it has the form of a state feedback plus an additional term that comes from the 
		 linear system~\req{po_DRE}.

		 Finally, the $p_0$ term in~\req{p2_DRE} is just the integral of the right hand side which depends 
		 on $p_1$ (and in turn $P$). This term plays no role in the optimal input, but it does appear in the 
		 value function, which is 
		 \begin{align} 
		 	 \smfm{ \bbm \xinitst ~~ 1 \ebm  \bbm P(0) & p_1(0) \\ p_1^*(0) & p_0(0) \ebm 	\bbm \xinit \\ 1 \ebm  }  	
			 	&\speq \xinitst P(0) \, \xinit + 2 \, p_1^*(0) \, \xinit + p_0(0), 
				&  \xinit = x(0)		.						
															\label{V_f_init.eq} 		
		 \end{align} 
		Note that this is an affine-quadratic function of the initial state $\xinit$. 
	
		\begin{remarc}							\label{H_R_init.remark}
			Even if $Q\geq 0$, the DRE~\req{Ph_eq} has a $\Qba$ matrix which is not guaranteed to 
			be positive semidefinite (this depends on also $q_1,q_0$). However, the block-triangular
			structure 
			of  $\Aba$ and $\Bba$ avoids this issue. We see from~\req{P1_DRE} that
			 $Q\geq 0$ and $R>0$ are sufficient to guarantee existence in that DRE.  
			Consequently solutions to~\req{po_DRE}-\req{p2_DRE} also exist over $[0,\sT]$ 
			since~\req{po_DRE} is a linear system 
			for $p_1$, and $p_0$ is simply the integral of the right hand side in~\req{p2_DRE}. 
			Thus solutions $\Pba$ to~\req{Ph_eq} exist over $[0,\sT]$ if $Q\geq 0$ and $R>0$. 
		\end{remarc}

	\subsection{Affine-Quadratic Problems with Final Conditions} 				\label{AQ_fin.sec}
	
		The DRE for the  homogenized problem~\req{quad_embed} with {\em final} conditions is given by~\req{DRE_fin}. 
		Label 
	its solution as $\Sba$ and partition it conformably with the partitions in~\req{A_B_part} as follows 
	\[
		\dot{\Sba} = -\Aba^*\Sba - \Sba \Aba - \Sba \Bba R^{-1} \Bba^* \Sba + \Qba , 
		\hstm \hstm \Sba = \smfm{ \bbm S & s_1 \\ s_1^* & s_0 \ebm } , 
		 \hstm \Sba(0) = \smfm{ \bbm \sfX & 0 \\ 0 & 0 \ebm } .
	\]
	Partitioning this DRE conformably with the partitioning of $\Aba, \Bba, \Qba$ gives	
		\begin{align*} 
			\hspace{-2em}
			{\arraycolsep=3pt
			 \smfm{ \bbm \dot{S} & \dot{s}_1 \\ \dot{s}_1^* & \dot{s}_0 \ebm } }
			=\,& {\arraycolsep=3pt
				\smfm{-\bbm A^* & 0 \\ 0 & 0 \ebm } 
				 \smfm{ \bbm S & s_1 \\ s_1^* & s_0 \ebm } - 
				 \smfm{ \bbm S & s_1 \\ s_1^* & s_0 \ebm }
				  \smfm{\bbm A & 0 \\ 0 & 0 \ebm } 
				  -
				  \smfm{ \bbm S & s_1 \\ s_1^* & s_0 \ebm } 
				  \smfm{ \bbm B \\ 0 \ebm } R^{\sm 1} \smfm{ \bbm B^* & 0 \ebm } 
				  \smfm{ \bbm S & s_1 \\ s_1^* & s_0 \ebm } 
				  + 
				  \smfm{\bbm Q &  q_1 \\ q_1^* & q_0 \ebm} 	}		\\
			=\,& 
				{\arraycolsep=3pt
				\smfm{-\bbm A^*S & A^*s_1 \\ 0 & 0 \ebm } 
				 - 
				  \smfm{\bbm SA & 0 \\ s_1^*A  & 0 \ebm } 
				  -
				  \smfm{ \bbm SBR^{\sm1}B^* S& SBR^{\sm1} B^* s_1  \\ 
				  			s_1^*BR^{\sm1}B^* S & s_1^*BR^{\sm1}B^* s_1 \ebm    } 
				  + 
				  \smfm{\bbm Q &  q_1 \\ q_1^* & q_0 \ebm} 	}.
		\end{align*} 
		This again leads to three equations for the (1,1), (1,2), and (2,2) blocks respectively 
		\begin{mytcolorbox}		\vspace{-1.3em}
		\begin{align} 
			 \dot{S} \speq& -A^* S - S A - S BR^{-1} B^* S + Q, 
				& S(0) &= \sfX,  							\label{P1_DRE_F.eq}	\\ 
			 \dot{s}_1 \speq& -A^* s_1 - S BR^{-1} B^* s_1 + q_1 , 
														\nonumber 		\\
					  \speq& -(A + BF )^*\,   s_1 + q_1 , 
					  \hstm F := R^{-1}B^* S ,
				& s_1(0) & = 0, 							\label{po_DRE_F.eq}	\\
			\dot{s}_0 \speq& - s_1^*BR^{-1}B^* s_1 + q_0 , 
				& s_0(0) &= 0, 							\label{p2_DRE_F.eq}
		\end{align} 
		\end{mytcolorbox}
	
		In contrast to the $\LQRi$ problem differential equations~\req{P1_DRE}-\req{p2_DRE}, 
		these equations can be solved 
		{\em forward} in time since initial conditions are given. We will see in the estimation problem 
		that the linear system~\req{po_DRE_F} will produce (after a change of variables)  the 
		(causal) observer structure of the Kalman filter.

		The optimal input for this problem obtained from~\req{DRE_fin} is given by 
		 \begin{align} 
		 	u =  R^{\sm1} \Bba^* \Sba \, \xi 
				\speq  R^{\sm1} \bbm B^* & 0 \ebm 
					\smfm{ \bbm S & s_1 \\ s_1^* & s_0 \ebm } 
					\smfm{\bbm x \\ 1 \ebm} 
				&\speq R^{\sm1} B^* \lb S \, x + s_1 \rb 			\nonumber		\\
				 &\speq {R^{\sm1} B^*  S \, x} 
							+  {R^{\sm1} B^* s_1}   , 
															 \label{u_opt_s.eq}
		 \end{align} 	
		while the optimal performance is a quadratic form on the (final) homogenized coordinates
		 \begin{align} 
		 	 \smfm{ \bbm \xfinst ~~ 1 \ebm  \bbm S(\sT) & s_1(\sT) \\ s_1^*(\sT) & s_0(\sT) \ebm 	\bbm \xfin \\ 1 \ebm  }  	
			 	&\speq \xfinst S(\sT) \, \xfin + 2 \, s_1^*(\sT) \, \xfin + s_0(\sT), 
				& \xfin  = x(\sT). 	
															\label{V_f_fin.eq} 	
		 \end{align} 
		Note again that this is an affine-quadratic functional on the final state $\xfin$. 


	In the next section we apply the initial-value affine-quadratic problem solution to the well-known 
	LQ-tracking problem. In the following Section~\ref{st_est.sec}, we apply the final-value affine-quadratic problem 
	to the state estimation problem. 
	
	\begin{remarc} 
	 	As in remark~\ref{H_R_init.remark}, the existence conditions for~\req{P1_DRE_F} are $Q\geq0$ 
		and $R>0$. 
		However, when this is applied to the estimation problem in Section~\ref{st_est.sec}, we will  require 
		$S(t)>0$.  The additional  condition of either $(Q,A)$ observable guarantees this
		 for $t\in(0,\sT]$,  or alternatively  $\sfX>0$ guarantees this for $t\in[0,\sT]$. 
	\end{remarc} 
	
	\section{Linear-Quadratic Tracking} 					\label{tracking.sec}
	
		The standard Linear-Quadratic tracking (LQ-tracking) problem (also known as the servomechanism 
		problem) involves linear dynamics with a quadratic tracking objective 
		as follows\footnote{For simplicity of exposition, a final state cost is not included here.}  
		\be 
			\begin{aligned} 
			\xd =& A \, x + B \, u , \hstm x(0) = \xinit , 		\\ 
			J :=& \smint{0}{T} \lb  \big( r - Cx \big)^*M \, \big( r - Cx \big) + u^* R \, u \rb dt ,
			\hstm M\geq 0, ~R>0, 
			\end{aligned} 
		  \label{LQ_track_form.eq}
		\ee
		where $Cx$ is considered as an output that should ``track'' a given reference signal 
		$r$. 

		The problem~\req{LQ_track_form}  is of the form~\req{qla_prob} since the objective is 
		affine-quadratic 
		\[
			J = \smint{0}{T} \Big( 
				x^* \underbrace{C^*MC}_{Q} \, x - 
				\underbrace{r^*MC}_{q_1^*}  \,  x - 
				x^* \underbrace{ C^*Mr}_{q_1}  + 
				\underbrace{r^*M\, r}_{q_0}  ~+~ u^* R \, u \Big)~ dt
		\]
		In homogeneous coordinates, this becomes the following initial-state LQR problem 
		\be
			\begin{aligned} 
			\smfm{\bbm \xd \\ \dot{\alpha} \ebm }
			=& \smfm{ \bbm A & 0 \\ 0 & 0 \ebm \bbm x \\ \alpha \ebm 
				+ \bbm B \\ 0 \ebm u }, 
				\hspace{10em}  \smfm{ \bbm x(0) \\ \alpha(0) \ebm = \bbm \xinit \\ 1 \ebm} , 		\\
			J =& \smint{0}{T}\!\! \lb  \smfm{ \bbm x \\ \alpha \ebm^*
					\bbm C^*MC & \sm C^*M r \\ \sm r^*MC & r^*M\, r \ebm 
						\bbm x \\ \alpha \ebm} + u^* R \, u \rb \,  dt  .
			\end{aligned} 
		 \label{LQT_reform_LQR.eq} 
		\ee
		 Note that 
		the state cost above includes the tracking signal $r$, so even if the original  LQ-tracking 
		problem is time invariant, its reformulation~\req{LQT_reform_LQR}
		 as an LQR problem has a time-varying state cost (a time-varying $Q$-matrix). 

		Equations~\req{P1_DRE}-\req{p2_DRE} in this case become 
		\begin{align} 
			- \dot{P} \speq& A^* P + P A - P BR^{\sm1} B^* P + C^*MC, 
				& P(\sT) &= 0, 											\label{P1_DRE_LQT.eq}	\\ 
			- \dot{p}_1 \speq& (A+BK)^* \, p_1  - C^*M \, r , 
				\hstm K := -R^{\sm1} B^* P,
				& p_1(\sT) & = 0, 											\label{po_DRE_LQT.eq}	\\
			- \dot{p}_0 \speq& - p_1^*BR^{\sm1}B^* p_1 + r^*M\, r , 
				& p_0(\sT) &= 0. 											\label{p2_DRE_LQT.eq}
		\end{align} 
		Note that Equation~\req{P1_DRE_LQT} is a DRE associated with an LQR 
		problem with the  matrices $(A,B,C^*MC,R)$. 
		It is time invariant if the original LQ-tracking problem is time invariant. 
		
		Equation~\req{po_DRE_LQT} is a linear system for the vector signal $p_1$, driven by the reference signal $r$, 
		and evolving anti-causally back from the final condition 
		\[
			\dot{p}_1(t) \speq -(A+BK)^* \, p_1(t) + C^*M \, r(t), 
			\hstm \hstm p_1(\sT) = 0, 
			\hstm \hstm K := -R^{\sm1}B^* P . 
		\]
		The optimal control $u$ is given by the formula~\req{u_opt}  as
		 \be
		 	u  
				\speq \underbrace{-R^{\sm1} B^*  P \, x}_{\mbox{\fs feedback}} 
							-  \underbrace{R^{\sm1} B^* \, p_1}_{\mbox{\fs feedforward}}   
		   \label{LQT_sol.eq}
		 \ee
		 This, together with the differential equations~\req{P1_DRE_LQT}-\req{po_DRE_LQT} 
		  is the well-known solution to the LQ-tracking problem 
		 (see e.g.~\cite[Sec. 5.2]{sage1977optimum} or~\cite{kirk1970optimal}). We list them 
		 all together next for later reference and comparison. 
		 \begin{mytcolorboxtitle}{Optimal LQ-Tracking Control} 	\vspace{-1em}
		 \be
		\begin{aligned} 
			- \dot{P} \speq& \, A^* P + P A - P BR^{\sm1} B^* P + C^*MC, 
				& P(\sT) &= 0, 												\\ 
			- \dot{p}_1 \speq& \, (A+BK)^* \, p_1  - C^*M \, r , 
				\hstm K := -R^{\sm1} B^* P,
				& p_1(\sT) & = 0, 										\\
			u \speq& -R^{-1} B^* \lb P x  +  p_1 \romn \rb  , 
			\hstm \xd \speq A \, x + B \, u, 		& x(0) &= \xinit. 
		\end{aligned} 
		\label{P1_DRE_LQT_box.eq}
		\ee
		 \end{mytcolorboxtitle} 
		 
		 \begin{remarc} 
		 	The conditions $M\geq 0$ and $R>0$ in~\req{LQ_track_form}  guarantee 
			$C^*MC\geq0$ in~\req{P1_DRE_LQT}
			and thus the existence of a solution $P(t)\geq 0$ over $[0,\sT]$. 
		 \end{remarc} 
		 
		 \begin{remarc}
		Sometimes an equivalent version of the solution is written using the negative of 
		 $p_1$ in~\req{po_DRE_LQT}. If we define $g(t) :=-p_1(t)$, then the differential equation for $g$ is 
		 \be
		 	\dot{g} = -(A+BK)^*  g - C^* M \, r , 
			\hstm \hstm g(\sT) = 0, 						\label{f_antic.eq}
		 \ee
		 and the control input can then be written as a function of the difference 
		 \[
		 	u \speq -R^{\sm1} B^* \lb P \, x - g \rb, 
		 \]
		 i.e. in this form the optimal control $u$ is a feedback on the {\em difference} between the 
		 state and the output $g$ of the anticausal linear system~\req{f_antic}. 
		\end{remarc} 

	\section{State Estimation} 			\label{st_est.sec}

	As described in the introduction,  the LSSE problem is 
	to find the signals $(\xh,\wh)$ that minimize the following cost $J$ subject to the
	 dynamic constraints 
	\be
		\begin{aligned}
			\dot{x} &= A \, x + w  ,
		\end{aligned}
		\hstm  
		\begin{aligned}
		J(x,w)  &=\!  \smint{0}{T} \!\!\lb  (y - Cx)^* \sfV\,  (y-Cx) + w^*\sfW \, w \romn \rb d\tau  
								+ \| x(0) \|_{\sfX}^2 , 		\\ 
					\sfV &> 0, ~ \sfW > 0, ~ (C,A)~\mbox{observable}, 
		\end{aligned} 
		\label{est_form_one.eq}
	\ee
	In homogeneous coordinates, this problem becomes 
	\be
	\begin{aligned} 
		\smfm{
		\bbm \dot{x} \\ \dot{\alpha} \ebm} &= 
		\smfm{ \bbm A & 0 \\ 0 & 0 \ebm \bbm  x \\ \alpha \ebm  \!+\! \bbm I \\ 0 \ebm  w} ,	
		\hstm \hstm    \alpha(\sT) = 1	,		\\ 
		J(x,w) 
			&= \! \smint{0}{T}  \Big( \smfm{ \bbm x \\ 1 \ebm^{\!*} \!
						\arraycolsep=2pt
						\bbm C^*\sfV C & \sm C^* \sfV y \\ \sm y^* \sfV C & y^*\sfV y \ebm 
									\bbm x \\ 1 \ebm  }  + w^*  \sfW w \Big) d\tau  
					+ \smfm{ \left\|  \bbm x(0) \\ 1 \ebm \right\|_{\sfXba}^2 }.  
	\end{aligned} 
	\label{est_form.eq}
	\ee
	This problem appears  similar to the affine-quadratic problems of Section~\ref{AQ_fin.sec} with 
	$B\, u$ replaced by $w$. 
	However, here neither initial $x(0)$ nor  final $x(\sT)$  states are   
	specified. In fact, finding the entire 
	state trajectory  (as well as the ``optimal'' estimate of the 
	process disturbance $w$) over $[0,\sT]$ is the task. 
	
	One way to approach this difficulty is to treat the problem~\req{est_form} as if the final state 
	is known, use  the final  condition LQR formulation for which we know the optimal 
	cost is given by~\req{V_f_fin}. The next step would be to optimize 
	the optimal cost further with respect to the unknown portion 
	 $x(\sT)=\xfin$ of the final state. This should yield the 
	answer to~\req{est_form} since it amounts to optimizing over all consistent state trajectories with unknown 
	final conditions. 
	
	Specifically, consider the problem 
	\be
	\begin{aligned} 
		\smfm{
		\bbm \dot{x} \\ \dot{\alpha} \ebm} &= 
		\smfm{ \bbm A & 0 \\ 0 & 0 \ebm \bbm  x \\ \alpha \ebm  \!+\! \bbm I \\ 0 \ebm  w} ,	
		    	\hstm \hstm \hstm 
			\smfm{\bbm x(\sT) \\ \alpha(\sT) \ebm = \bbm \xfin \\ 1 \ebm ,}		\\ 
		J(x,w) 
			&= \! \smint{0}{T}  \Big( \smfm{ \bbm x \\ 1 \ebm^{\!*} \!
						\arraycolsep=2pt
						\bbm C^*\sfV C & \sm C^* \sfV y \\ \sm y^* \sfV C & y^*\sfV y \ebm 
									\bbm x \\ 1 \ebm  }  + w^*  \sfW w \Big) d\tau  
					+ \smfm{ \left\|  \bbm x(0) \\ 1 \ebm \right\|_{\sfXba}^2 }.  
	\end{aligned} 
	\label{est_form_xfin.eq}
	\ee
	This is a problem of the form~\req{quad_embed} with a known final state. 
	The optimal solution to such problems  is given by 
	Equations~\req{P1_DRE_F}-\req{p2_DRE_F}, which for this case  become 
		\begin{mytcolorbox}		\vspace{-1.3em}
		\begin{align} 
			 \dot{S} \speq& -A^*S - S A - S \, \sfW^{\sm1}  S + C^*\sfV C, 
				& S(0) &= \sfX, 											\label{P1_DRE_K.eq}	\\ 
			 \dot{s}_1 \speq& -(A+\sfW^{\sm1}S)^* \, s_1  - C^*\sfV \, y , 
				& s_1(0) & = 0, 											\label{po_DRE_K.eq}	\\
			 \dot{s}_0 \speq& - s_1^*\sfW^{\sm1} s_1 + y^*\sfV y , 
				& s_0(0) &= 0. 											\label{p2_DRE_K.eq}
		\end{align} 
		\end{mytcolorbox}		
		
	The optimal value of the problem~\req{est_form_xfin} 
	constrained by $x(\sT)=\xfin$ is given by~\req{V_f_fin}  as 
	\be
		\inf_{x,w,~x(\ssT)=\xfin} J(x,w) \speq \xfinst  S(\sT) \, \xfin + 2 \, s_1^*(\sT) \, \xfin + s_0(\sT). 
	  \label{aq_Ss1.eq}
	\ee
	If we optimize this further with respect to $\xfin$, we actually obtain the optimal estimate 
	$\xh(\sT)$ at time $\sT$. 
	The  affine-quadratic functional~\req{aq_Ss1} is minimized with respect to $\xfin$ by 
	\be
		\arg\min_{\xfin} 	\lb 	\xfinst  S(\sT) \, \xfin + 2 \, s_1^*(\sT) \, \xfin + s_0(\sT)  \romn \rb 
		\speq - S^{-1}(\sT) \, s_1(\sT)  \speq \xh(\sT) .  				\label{xh_s1.eq}
	\ee
	Thus the optimal estimate $\xh(\sT)$ is obtained from $S(\sT)$ and $s_1(\sT)$.

	Now one way to obtain the optimal estimate $\xh(\sT)$ is to run 
	Equations~\req{P1_DRE_K}-\req{po_DRE_K} simultaneously 
	forward in time to compute  $S(\sT)$ and $s_1(\sT)$, and then obtain $\xh(\sT)$ from~\req{xh_s1}.  
	However, we may want to directly find a differential equation (differentiating with respect 
	to $\sT$) that $\xh(\sT)$ satisfies and run that 
	instead of the equation for $s_1$. Indeed, the differential equation for $\xh(\sT)$ can be derived 
	from~\req{xh_s1} and~\req{P1_DRE_K}-\req{po_DRE_K} as follows
	(all derivatives below 
	are with respect to $\sT$, e.g. $\dot{\xh} := \tfrac{d}{d\ssT} \xh(\sT) $) 
	\begin{align} 
		\dot{\xh} 
			&= S^{-1} \dot{S} \, S^{-1} \, s_1 - S^{-1} \, \dot{s}_1 
							\tag{\fs using~\req{xh_s1} and the product rule}	\nonumber	\\
			&= S^{\sm1} \lb -A^*S - S A - S\sfW^{\sm1}  S + C^*\sfV C \rb  S^{\sm1} \, s_1 
				+ S^{\sm1} \,  \lb (A+\sfW^{\sm1}S)^* \, s_1  +  C^*\sfV \, y   \rb 			\nonumber	\\
			&= \lb - \cancel{S^{\sm1} A^*} -  AS^{\sm1} - 
				\cancel{\sfW^{\sm1} }  + S^{\sm1}C^*\sfV CS^{\sm1} \rb  s_1  
				+  \cancel{S^{\sm1}A^* \, s_1} + \cancel{\sfW^{\sm1} \, s_1} 
				+  S^{\sm1} C^*\sfV \, y   								 		\nonumber	\\
			&=   -  A~ S^{-1} s_1    + S^{-1}C^*\sfV C ~  S^{-1}    s_1  
				 +  S^{-1} C^*\sfV \, y    										\nonumber	\\
			&=     A \, \xh    -  S^{-1}C^*\sfV C\, \xh  
				 +  S^{-1} C^*\sfV \, y    										\nonumber	\\
			\hspace{-2em}
			\highcolbox{
			\Rightarrow \raisebox{-.25em}{~}\hsom \dot{\xh}  }
			&\highcolbox{=     \lb A     -  LC \rb  \xh   +  L \, y			 
				 	\speq A \, \xh + L\lb y - C\xh \rb , 
					\hstm \hstm 		L := S^{-1} C^* \sfV    \raisebox{.93em}{~}	.}	
																\label{xh_obs_der.eq} 
	\end{align} 
	Thus the dynamics of $\xh$ are those of an observer for the system $\xd = Ax$, $y=Cx$, 
	with an optimal observer gain $L$ obtained from the solution of the forward DRE~\req{P1_DRE_K}. 
	Equations~\req{P1_DRE_K} and~\req{xh_obs_der} are the ``information filter'' form of the 
	Kalman filter. This terminology and the relations to the stochastic version of the Kalman 
	filter (in terms of covariance matrices) are detailed in Section~\ref{KF_cov.sec}. 
	Note that there is no dependence on $s_1$ in~\req{xh_obs_der} since $s_1$ has been 
	eliminated using~\req{xh_s1}. The next subsection on the ``smoothing'' problem clarifies this issue.

	\begin{remarc} 
		Sufficient   conditions for existence in the DRE~\req{P1_DRE_K} are $\sfV\geq 0$ and $\sfW>0$. However, 
		the formula~\req{xh_obs_der} for the observer gain $L$ requires $S$ to be invertible. 
		Either $\sfX>0$ or observability of  $(C^*\sfV C, A)$  guarantees $S(t)>0$. 
		The problem~\req{est_form_one}  was stated with the assumption of observability of $(C,A)$, 
		and  since $\sfV>0$ then 
		 $(\sfV^{\frac{1}{2}}C,A)$ is also observable, which is equivalent to 
		$(C^*\sfV C, A)$ observable. 
%
	\end{remarc} 
	
	A less stringent condition can also be stated for $S(t)>0$. See Remark~\ref{Nus.remarc} in the next section.

	\subsection{Smoothing vs. Filtering} 							\label{smooth.sec}
	
	The estimate $\xh$  in~\req{xh_obs_der} is the causal  estimate of $x(\sT)$, i.e.  
	based on observations 
	$\yoTc$, or equivalently the estimate of $x(t)$ based on 
	$\yotc$ (just relabeling $\sT$ as $t$). This is sometimes written as $\xh\big(t\,  \pmb{|} {\scz [0,t]}\big)$, 
	and referred to as the ``filtered estimate''. 
	Another quantity of interest is the so-called ``smoothed'' estimate, which is the estimate of 
	$x(t)$ (for $t\in[0,\sT]$) based on past and future measurements, i.e. based on 
	$\yoTc$. 
	This is  sometimes denoted by $\xh\big(t\,  \pmb{|} {\scz [0,\sT]}\big)$. This notation of course 
	originates from conditioning notation in probability. 
	
	We will investigate the relations between the filtered and smoothed estimate, but using 
	the following clearer notation (in the deterministic setting) 
	\begin{align*} 
		\xh_\rmf (t) :=&\,  \mbox{estimate of $x(t)$ given $\yotc := \lcb y(\tau); ~ \tau\in[0,t] \rcb$}	
				 \tag{filtered estimate} 	\\
		\xh_\rms (t) :=&\,  \mbox{estimate of $x(t)$ given $\yoTc := \lcb y(\tau); ~ \tau\in[0,\sT] \rcb$}
				 \tag{smoothed estimate} 
	\end{align*} 
	Clearly $\xh_\rmf(\sT) = \xh_\rms(\sT)$ since they are based on the same  signal $\yoTc$. However, 
	for $t<\sT$, $\xh_\rms(t)$ should be better than $\xh_\rmf(t)$ since the former uses all of $\yoTc$, 
	while the latter only uses the $y_{\sss [0,t]}$ portion. 
	
	Note that the problem~\req{est_form_one} as stated should produce the optimal 
	input  $\lcb \wh(t) ; ~t\in[0,\sT] \rcb$, as well as the optimal $\lcb \xh_\rms(t) ; ~t\in[0,\sT] \rcb$, 
	i.e. the smoothed estimate, 
	since that is the estimate for each $t\in[0,\sT]$ based on the signal 
	$\lcb y(t) ; ~t\in[0,\sT] \rcb$. Then in~\req{xh_s1} we only used the estimate at the end point, 
	i.e. $\xh_\rmf(\sT) = \xh_\rms(\sT)$ in our current notation. Also recall  that the differential 
	equation~\req{xh_obs_der} (for $\xh_\rmf$) is derived by differentiating with respect to the end time $\sT$. 
	
	There is an interesting dynamical structure in the smoothed estimate which connects it more closely 
	with the LQ-tracking problem. 
	The smoothed estimate $\lcb \xh_\rms(t); ~ t\in [0,\sT] \rcb$ is the solution to  
	the estimation problem~\req{est_form_xfin}. Since at the end time we have 
	$\xh_\rms(\sT)=\xh_\rmf(\sT)$, and if $\xh_\rmf(\sT)$ is known (from the filter~\req{xh_obs_der} say), 
	then  $\lcb \xh_\rms(t); ~ t\in [0,\sT] \rcb$ is the solution to the 
	problem~\req{est_form_xfin} with prescribed final state. 
	This ``input design'' problem  is 
	solved by~\req{P1_DRE_K}-\req{po_DRE_K}, 
	and the optimal input $\wh$ is given 
	from~\req{u_opt_s} as (with $B=I$ and $R=\sfW$) 
	\be
		\wh(t)  \speq \sfW^{-1} S(t) \, \xh_\rms(t)  + \sfW^{-1} \, s_1(t) , 
			\hstm t\in[0,\sT]. 
	  \label{wh_diffeq.eq} 
	\ee	
	Note that $(\wh,\xh_\rms)$ is the solution to the intermediate problem~\req{est_form_one} 
	over $[0,\sT]$. This last formula gives $\wh$ as a ``state feedback'' on $\xh_\rms$ (not $\xh_\rmf$), 
	plus a ``feedforward'' term $\sfW^{-1} s_1$. On the other hand, 
	 we know from~\req{xh_s1} that $\xh_\rmf$ and $s_1(t)$ are 
	related by 
	\[
		\xh_\rmf(t) \speq -S^{-1}(t) \, s_1(t) 
		\hstm \Leftrightarrow \hstm 
		s_1(t) \speq -S(t) \, \xh_\rmf(t) . 
	\]
	Substituting this into~\req{wh_diffeq} gives
	\be
		\wh(t) \speq \sfW^{-1} S(t) \lb \xh_\rms(t) - \xh_\rmf(t) \romn \rb , 		\label{wh_as_xh.eq}
	\ee
	i.e. $\wh$ is proportional to the difference between the filtered and smoothed estimates. Note that 
	at the boundary $t=\sT$, we have $\wh(\sT) =0$. 
	
	Using the relations above, a backwards differential equation can be written for the smoother.
	Indeed $\xh_\rms$ and $\wh$ satisfy the system equation $\dot{\xh}_\rms= A \, \xh_\rms + \wh$, 
	which after 
	incorporating~\req{wh_as_xh} yields 
	\begin{mytcolorbox}\vspace{-1em}
	\be
		\dot{\xh}_\rms(t)  \speq A \, \xh_\rms(t) + \sfW^{-1} S(t) \lb \xh_\rms(t) - \xh_\rmf(t) \romn \rb , 
		\hstm t\in[0,\sT], \hstm \xh_\rms(\sT) = \xh_\rmf(\sT). 
	\ee
	\end{mytcolorbox} 
	\noindent
	This is precisely the Rauch-Tung-Striebel smoother~\cite{rauch1965maximum}  originally derived in the 
	context of maximum likelihood  Gaussian
	estimation (see also~\cite[Sec. 5.2]{gelb1974applied}). 
	It is interesting  here that it falls out of this deterministic formulation 
	rather effortlessly. Note that for any $t<\sT$, this estimator is not a causal system since it requires 
	the terminal condition $\xh_\rms(\sT)=\xh_\rmf(\sT)$, which is obtained from the whole data record  
	 $\yoTc$ over $[0,\sT]$. 
	
	We can also rewrite the smoother equations with $s_1$ as the input as follows
	\begin{mytcolorboxtitle}{Least-Squares State Smoother}	\vspace{-1em}
		\be
		\begin{aligned} 
			 \dot{S} \speq& -A^*S - S A - S \, \sfW^{\sm1}  S + C^*\sfV C, 
				& S(0) &= \sfX, 											\\ 
			 \dot{s}_1 \speq& -(A+\sfW^{\sm1}S)^* \, s_1  - C^*\sfV \, y , 
				& s_1(0) & = 0, 											\\
			\wh \speq& \, \sfW^{-1} S \, \xh_\rms  + \sfW^{-1} \, s_1, 
				\hstm \dot{\xh}_\rms = A \, \xh_\rms + \wh , 
				& \xh_\rms(\sT) &= \xh_\rmf(\sT). 									
		\end{aligned} 						\label{P1_DRE_K_smooth.eq}
		\ee
	\end{mytcolorboxtitle}		
	\noindent
	where we simply used the formula~\req{wh_diffeq} for the optimal $\wh$, and the 
	system equation $ \dot{\xh}_\rms = A \, \xh_\rms + \wh $. This can now be compared 
	with the LQ-tracking controller as written in~\req{P1_DRE_LQT_box}. Here $\wh$ is the 
	control, and the state $\xh_\rms$ is the smoothed estimate. The Riccati equation 
	and the linear system driven by the external signal ($y$ here, compared to 
	 $r$ in~\req{P1_DRE_LQT_box}) both 
	have initial rather than final conditions as in~\req{P1_DRE_LQT_box}. The state equation 
	however for $\xh_\rms$ has a final condition rather than an initial condition for 
	the state equation in~\req{P1_DRE_LQT_box}. 
	
	It appears that Equations~\req{P1_DRE_K_smooth} 
	are equivalent (in form) to~\req{P1_DRE_LQT_box} with a time reversal. This is indeed the case 
	if we reverse time in~\req{P1_DRE_LQT_box} by defining $S(t) := P(\sT-t)$, $s_1(t):=p_1(\sT-t)$, 
	$\xh_\rms(t):= x(\sT-t)$,  $\wh(t):= u(\sT-t)$ and $y(t):= r(\sT-t)$, where $P,p_1,u,x,r$ satisfy~\req{P1_DRE_LQT_box}. 
	First the differential equation relating $\xh_\rms, \wh$ becomes 
	\[
		\xd = A \, x + B \, u 
		\hstm \Rightarrow \hstm 
		-\dot{\xh}_\rms = A \, \xh_\rms + B \, \wh 
		\hstm \Rightarrow \hstm 
		\dot{\xh}_\rms = -A \, \xh_\rms - B \, \wh . 
	\]
	In this equation, let  $B = -I$ and relabel $F:=-A$, then  Equations~\req{P1_DRE_LQT_box} become 
		 \be
		\begin{aligned} 
			 \dot{S} \speq& - F^* S - S F - S R^{\sm1}  S + C^*MC, 
				& S(0) &= 0, 												\\ 
			 \dot{s}_1 \speq& \lb -F^* - SR^{-1} \rb   s_1  - C^*M \, y , 
				& s_1(0) & = 0, 											\\
			\wh \speq& R^{-1}  \lb S \, \xh_\rms  +  s_1 \romn \rb  , 
			\hstm \dot{\xh}_\rms \speq F \, \xh_\rms +  \wh, 	\hstm 	& \xh_\rms(\sT) &= \xinit. 
		\end{aligned} 
		\label{P1_DRE_LQT_reverse.eq}
		\ee
	These equations are exactly the  equations of the least-squares smoother above 
	 if we further identify $R,M,\xinit$ 
	here with $\sfW,\sfV,\xh_\rmf(\sT)$
	respectively in~\req{P1_DRE_K_smooth}. The reason $S(0)=0$ here rather than $S(0)=\sfX$ 
	as in~\req{P1_DRE_K_smooth} 
	is that the problem~\req{P1_DRE_LQT_box} was derived without a final state penalty. Had the 
	final state penalty $x^*\!(\sT) \, \sfX \, x(\sT)$ been there, we would have $S(0) = \sfX$ in~\req{P1_DRE_LQT_reverse}
	as expected.

	\section{Relations to the Stochastic  Kalman-Bucy Filter} 			\label{KF_cov.sec}
	
	The stochastic version of the Kalman filter problem is formulated for the following  dynamics 
	\[
		\begin{aligned} 
			\xd(t)  & = A \, x(t) + w(t) , 		\\ 
			y(t) &= C \, x(t) + v(t) , 
		\end{aligned} 
	\]
	where $w$ and $v$ are zero-mean, mutually uncorrelated, 
	 white noise  second-order processes with the following 
	statistics 
	\begin{align*} 
			\expct{ w(t) \, w^{\!*}\! (\tau) } &= \Sigma_w\,  \delta(t- \tau), 
					& \expct{w(t) \, v^{\!*}\!(\tau)} &= 0	,				\\ 
			\expct{ v(t) \, v^{\!*}\!(\tau)}  & = \Sigma_v\,  \delta(t- \tau) , 	
					& \expct{x(0) \, x^{\!*}\!(0) } &= \Sigma_\rmi ,
					& \expct{x(t) \, w^{\!*}\!(\tau) } & = 0, ~\tau \geq t, 
	\end{align*} 
	where $\Sigma_w$ and $\Sigma_v$ are the instantaneous covariance matrices of 
	 $w$ and $v$ respectively, while $\Sigma_\rmi$  is the covariance matrix 
	of the (assumed zero-mean) initial state. The objective is to find the estimate $\xh$ which minimizes the 
	variance    
	\[
		 \expct{ \| e(t) \|^2} = 
		 \expct{ \| x(t) - \xh(t) \|^2} 
	\]
	of the estimation error $e := x-\xh$ at each time $t$. 
	
	As is well known, the solution of this problem is given by a DRE for the error covariance matrix $\Sigma_e$, 
	together with an observer for the dynamics of $\xh$ 
	\begin{mytcolorbox}	\vspace{-1em}
	\begin{align} 
		\dot{\Sigma}_e &= A\,  \Sigma_e + \Sigma_e\,   A^* - \Sigma_e\,  C^* \Sigma_v^{-1} C \, \Sigma_e
							+ \Sigma_w , 
					& \Sigma_e(0) =&\,  \Sigma_\rmi , 		\label{Sigma_e_RDE.eq} 	\\ 
		\dot{\xh} &= A \, \xh + L \lb y - C \, \xh \rb , 
					& L :=&\,  \Sigma_e C^* \Sigma_v^{-1}. 	\label{L_Sigma.eq} 
	\end{align} 
	\end{mytcolorbox}
	\noindent
	For comparison, the equations~\req{P1_DRE_K},~\req{xh_obs_der}  
	for $\xh$ from the deterministic problem  are
	\begin{mytcolorbox}	\vspace{-1em}
	\begin{align} 
		\dot{S} &= -A^*  S -  S\,    A - S\,  \sfW^{-1} S + C^*\sfV \, C  , 
					& S(0) =& \, \sfX , 					\label{S_RDE.eq} 			\\ 
		\dot{\xh} &= A \, \xh + L \lb y - C \, \xh \rb , 
					& L :=& \, S^{-1}  C^* \sfV. 				\label{L_S.eq}
	\end{align} 
	\end{mytcolorbox}

	The relation between the two is 
	$S(t) = \Sigma_e^{-1}(t)$.  Indeed, starting from
	the DRE~\req{S_RDE} for $S$, we can compute the differential equation for $S^{-1}$ as 
	\begin{align*} 
		\tfrac{d}{dt} S^{-1} &= - S^{-1} \dot{S}  \, S^{-1}  				\\ 
				& = -S^{-1} \lb 	 -A^*  S -  S\,    A - S\,  \sfW^{-1} S + C^*\sfV \, C  \rb S^{-1} 
							\mytag{using~\req{S_RDE} } 			\\
		\Rightarrow \hstm 
		\tfrac{d}{dt} S^{-1} 		&=  S^{-1} A^* + A \, S^{-1} + \sfW^{-1} - S^{-1} C^*\sfV C \, S^{-1} .
	\end{align*} 
	This last equation  is precisely~\req{Sigma_e_RDE} for $\Sigma_e$  if we identify 
	\[
		\Sigma_e = S^{-1}, 
		\hstm 
		\Sigma_\rmi = \sfX^{-1} , 
		\hstm 
		\Sigma_v = \sfV^{-1}, 
		\hstm 
		\Sigma_w = \sfW^{-1} . 
	\]
	Note also that  the two 
	observer gains $L$ in~\req{L_Sigma} and~\req{L_S} are equal as well 
	\[
		L \speq \Sigma_e C^* \Sigma_v^{-1} \speq S^{-1} C^* \sfV . 
	\]
	
	The matrix $S(t)=\Sigma_e^{-1}(t)$ is sometimes referred to as the ``information matrix''. 
	See below for further comments on this interpretation. 
	The standard interpretations of $\Sigma_w$ and $\Sigma_v$ are inverted when considering 
	the weight matrices $\sfW$ and $\sfV$ as follows. 
	\begin{itemize} 
		\item When $\Sigma_w$ is chosen small, this means lower error in the dynamics 
			$\xd = A \, x$, i.e. higher trust in the dynamical model. This is the same as choosing 
			$\sfW=\Sigma_w^{-1}$ large, which in this case will force   $w$ in the cost~\req{MSE_cost}
			to be small. 
		\item When $\Sigma_v$ is chosen small, this means higher trust in the measurements. 
			This is equivalent to choosing $\sfV=\Sigma_v^{-1}$ large, which will force the term $v$ in 
			the cost~\req{MSE_cost} to be small. 
		\item Lower uncertainty in the initial state means small $\Sigma_\rmi$. Again, this corresponds
			to choosing $\sfX= \Sigma_\rmi^{-1}$ large to force $\xinit$ in~\req{MSE_cost} to be small.  
	
	\end{itemize} 
	
	\begin{remarc} 							\label{Nus.remarc}
		A less stringent condition for  positivity $S(t)>0$,  $t\in(0,\sT]$  can be stated. Let $\calN_{\overline{o}}$
		denote the unobservable subspace of $(C,A)$, and $\Nus{\sfX}$ the null space of $\sfX$. Then 
		$S(t)>0$,  $t\in(0,\sT]$ iff $\calN_{\overline{o}} \cap \Nus{\sfX}=0$. This condition holds 
		if either $\sfX>0$ or $(C,A)$ observable, but it can also hold more generally. For example, 
		the  stochastic 
		version of the estimator can be run with $\Sigma_\rmi>0$ and no observability conditions. 
		Since $\sfX = \Sigma_\rmi^{-1}$, then $\Sigma_\rmi >0 ~\Leftrightarrow~ \sfX>0$. 
		The condition $\calN_{\overline{o}} \cap \Nus{\sfX}=0$ has a nice interpretation; 
		prior information in state space  is needed in  directions the output cannot observe. 
	\end{remarc}

	\subsection{Interpretations of the Covariance $\Sigma_e$ and Certainty $S$ Matrices} 
	
	When the covariance matrix $\Sigma_e(t) > 0$ is positive, then $S(t) = \Sigma_e^{-1}(t)$ and is also 
	positive.  Such matrices have nice geometric 
	interpretations in terms of ellipsoids. We begin with the covariance matrix $\Sigma_e$. 
	
\begin{figure}[t]
		\centering
		\begin{subfigure}[t]{0.41\textwidth}
			\centering 
			\includegraphics[width=.9\textwidth]{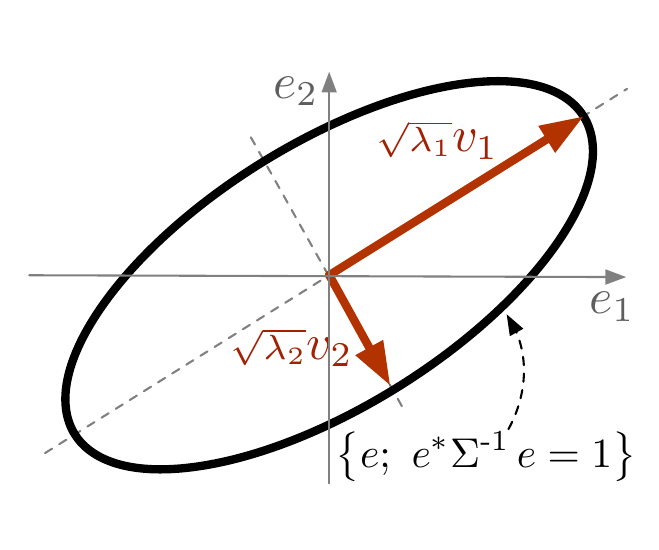} 
			
			\mysubcaption{The level set $\lcb e; ~ e^*\Sigma^{\sm1}_e e=1 \rcb$ for any matrix $\Sigma_e>0$
				is an ellipsoid whose principal axes align with the eigenvectors and eigenvalues 
				of $\Sigma_e$  as shown. Here $\lambda_1>\lambda_2$ are eigenvalues of $\Sigma_e$ 
				(not $\Sigma_e^{\sm1}$). 
				} 
		  \label{ellipsoids_eigs_LS.fig} 
		\end{subfigure}
		\hfill 
		\begin{subfigure}[t]{0.56\textwidth}
			\centering 
			\includegraphics[width=.8\textwidth]{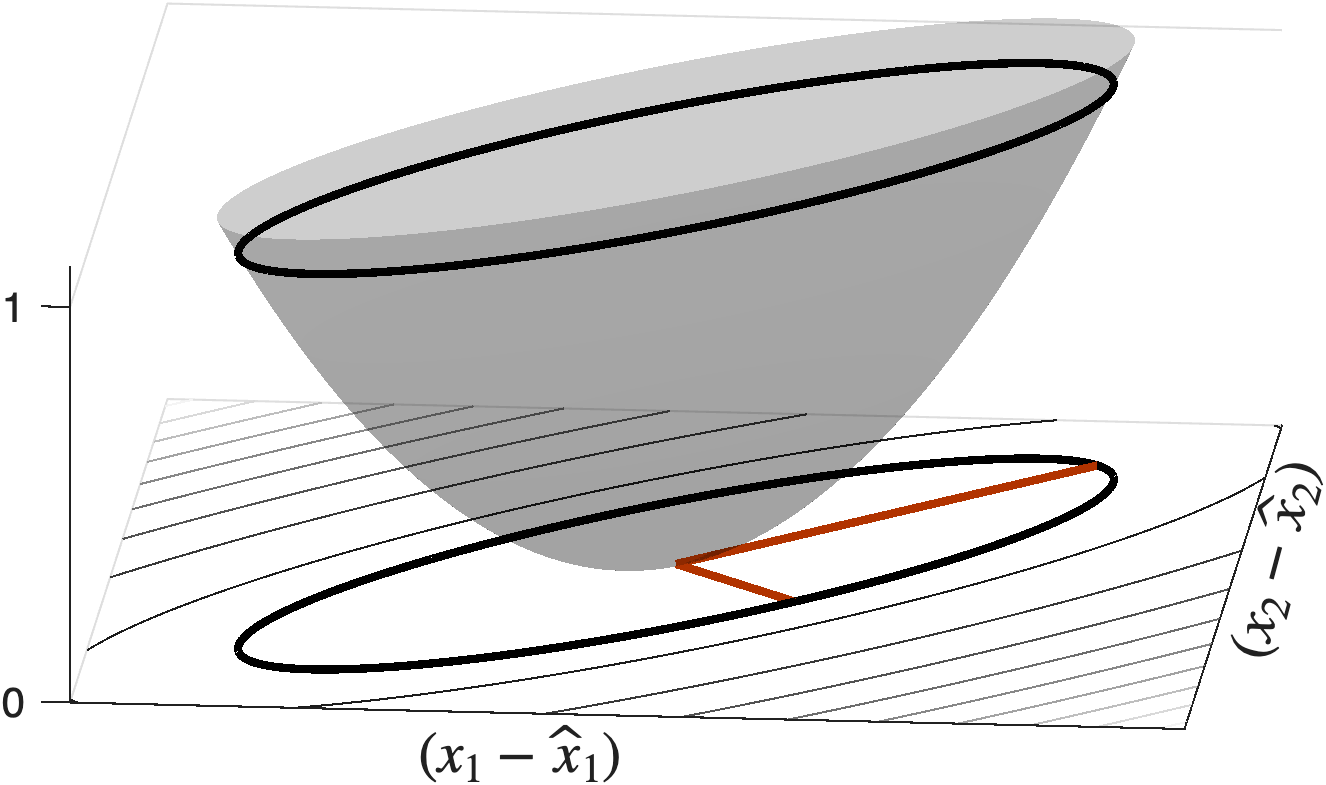} 
			
			\mysubcaption{The ``certainty functional'' 
				$(x\sm\xh)^*S \, (x\sm\xh) = (x\sm\xh)^*\Sigma^{\sm1} (x\sm\xh)$. Its  
				steepest ascent is in the direction of $v_2$, the eigenvector of $\Sigma$ with 
				the smallest eigenvalue, which corresponds to the largest eigenvalue of $S=\Sigma^{\sm1}$. 
				This is the direction of highest certainty, corresponding to the direction of lowest 
				variance in the stochastic interpretation. 
				} 
		  \label{S_plot.fig} 
		\end{subfigure}
		
		\bigskip 
		
		\begin{subfigure}[t]{0.9\textwidth}
			\centering 
			\includegraphics[width=.9\textwidth]{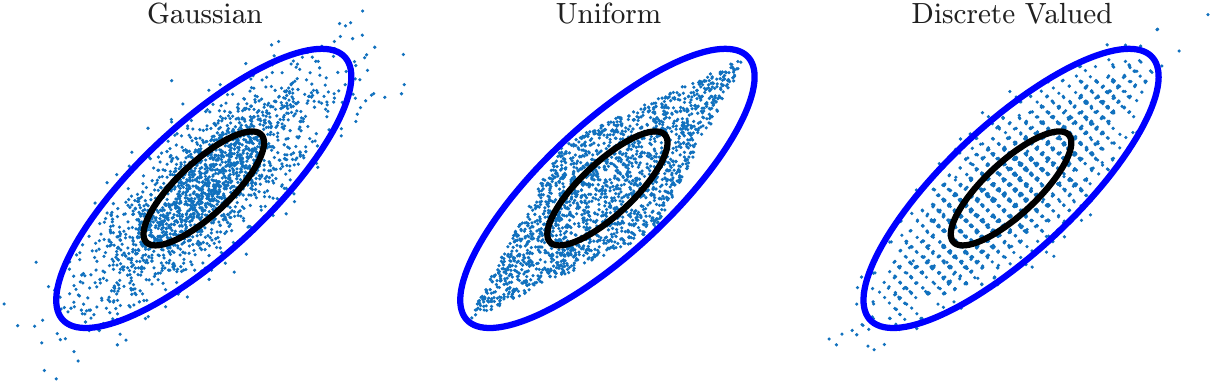} 
			
			\mysubcaption{Scatter plots of 3 random variables with different densities, normalized 
				to have the same covariance matrix $\Sigma$. The black ellipsoids 
				are the level sets $\lcb x; ~ x^*\Sigma^{\sm1} x = 1 \rcb$.  The blue ellipsoids
				are scaled to contain 95\% of the probability  mass for a Gaussian density. 
				} 
		  \label{rand_ellip.fig}
		\end{subfigure}	
		
		\mycaption{Geometric interpretation of  covariance $\Sigma$ and certainty $S=\Sigma^{-1}$ matrices, 
			which share the same eigenvectors, but have reciprocal eigenvalues.   Directions 
			of highest (lowest) variance correspond to directions of lowest (highest) certainty. 
			} 
	
	\end{figure}

	For  zero-mean Gaussian random vectors, 
		the covariance matrix completely determines the density 
	which is of the following form 	
	\[
	        	\rho(e)  \speq 
		 \tfrac{1}{\sqrt{(2\pi)^n \det(\Sigma_e)}} ~\exp \lb {\scriptstyle -\frac{1}{2}}~ e^* \Sigma_e^{\sm1} ~\!e \rb  .
	\]
	One way to visualize this density is through its level sets. The level sets of $\rho(.)$ are the 
	same as the level sets of the quadratic functional $e^* \Sigma_e^{\sm1} \, e$ (with different level 
	values). The level sets of this quadratic functional are ellipsoids with principal axes 
	aligned with the eigenvector directions. Let $\lcb \lambda_1, \ldots, \lambda_n \rcb$ be the 
	eigenvalues of $\Sigma_e$ (not $\Sigma_e^{\sm1}$)  arranged in descending order, with 
	 $\lcb v_1, \ldots, v_n\rcb$ the corresponding eigenvectors. 
	The ellipsoid  $\lcb x; ~ x^*\Sigma_e^{\sm1} \, x = 1\rcb$ has major axis with length 
	$2\sqrt{\lambda_1}$ in the direction $v_1$, while the minor axis has length $2\sqrt{\lambda_n}$ 
	in the direction $v_n$. See Figure~\ref{ellipsoids_eigs_LS.fig} for an illustration in two dimensions.

	For a Gaussian density, the fraction $p$ of the probability mass is contained in the level-set interior 
	\be
		\lcb e\in\R^n; ~ e^*\Sigma^{-1} \, e \leq F^{-1}_{\chi^2_n}(p)  \romn \rcb 
	  \label{prob_mass.eq}
	\ee
	where $F^{-1}_{\chi^2_n}$ is the inverse of the  cumulative density (the quantile function) of the 
	$\chi^2$ distribution with $n$ degrees of freedom. 
	Figure~\ref{rand_ellip.fig} shows the level sets in dimension 2 for three random vectors with 
	different densities. While the formula~\req{prob_mass} is exact only for Gaussian densities, it 
	is generally a reasonable approximation for some unimodal densities as seen in the figure. 
	Note how the major axis of the ellipsoid (corresponding to the largest eigenvalue $\lambda_1$ 
	of $\Sigma$) is  the direction of highest variance, 
	and vice versa for the minor axis. 
	
	Thus the eigenvalues/vectors of $\Sigma_e$ determine the variance of the estimation error in 
	various directions in state space. The variance of $e$ can be thought of as a proxy for 
	how much uncertainty there is in estimating that direction in state space. For most problems, 
	some directions in state space are easier to estimate than others. The eigenvalues/vectors 
	of $\Sigma_e$ quantify  this.

	In the 
	``information filter'' interpretation~\cite{ThrunLiuKollerNgGhahramaniDurrantWhyte2004SEIF} of 
	the Kalman filter, $\Sigma_e^{-1}(t)$ is called the \deffont{information matrix}, and 
	$\Sigma_e^{-1}(t)\,  \xh(t)$ is called the \deffont{information vector}. Note that the latter 
	is precisely $-s_1$ in the linear system~\req{po_DRE_K} since by~\req{xh_s1} 
	$ s_1(t) = - S(t) \, \xh(t) = -\Sigma_e^{-1}(t) \, \xh(t)$.
	
	The interpretation of $S(t)$ as a kind of ``information'' can also be given without resort to 
	probabilistic arguments. Recall that the optimal estimate $\xh$ is obtained 
	by optimizing~\req{aq_Ss1}  with respect to $\xfin$ 
	\be
		\xh \speq \arg\min_{\xfin} \lb  \xfinst  S \, \xfin + 2 \, s_1^* \, \xfin + s_0\rom \rb 
			\speq - S^{-1} s_1 .
	  \label{CTG_opt.eq}
	\ee
	Thus the difference between the cost-to-arrive (recall that this quantifies the  level  
	of uncertainty~\req{l_uncert} in the estimation problem)  evaluated at any ``guess'' $x$ compared to 
	it evaluated at the optimal vector $\xh$ is 
	\begin{align} 
		& 	 \lb x^* S \, x + 2\, s_1^* x + s_0 \rb	 
			- \lb \xh^* S \, \xh + 2\, s_1^* \xh + s_0 \rb			\nonumber	\\ 
		=&\,   \lb x^* S \, x - 2\, \xh^* S x \rb
			-  \lb \xh^* S \, \xh - 2\, \xh^* S \,  \xh \rb 
				\mytag{using $s_1^* = -\xh^*S^*$ from~\req{CTG_opt}}			\\
		=&\,  x^*S \, x - 2 \, \xh^* S \, x + \xh^* S \, \xh \speq 
			\lb x - \xh \rb^* \! S \lb x -  \xh \rb .				\label{certain_S.eq}
	\end{align} 
	With $S>0$, this is a quadratic, ``bowl-shaped'' functional of the difference between 
	the optimal estimate $\xh$ and any other vector $x$. For any given  direction $(x-\xh)$, 
	the steeper this functional 
	is, the more ``certain''  the optimal filter is of its estimate in that direction. In particular, the direction 
	of the eigenvector of $S$ of largest eigenvalue is the direction in which the filter is 
	most certain of its estimate, and  conversely the eigenvector associated with 
	the smallest eigenvalue is the direction of least certainty. 
	See Figure~\ref{S_plot.fig} for an illustration.

	Thus the matrix $S$ describes {\em how ``certain'' the optimal filter is of its state estimates 
	in various directions}. Here ``certainty'' in the sense of the curvature of the 
	functional~\req{certain_S} can be thought of as the deterministic counterpart of ``information'' 
	in the probabilistic setting. Note that the statements about eigenvectors of $\Sigma_e = S^{-1}$ 
	are reversed, e.g. the eigenvector of $\Sigma_e$ with largest eigenvalue represents the direction 
	in which the estimate has highest variance, i.e. highest uncertainty. This is precisely 
	the eigenvector of $S$ with lowest eigenvalue. 
	Thus in 
	 the deterministic formulation of the Kalman filter, it is perhaps better to refer to $S$ 
	as the ``certainty matrix'' rather than the ``information matrix''.

	\begin{remarc} 
		The stochastic version of the estimator requires $\Sigma_\rmi$ to initialize~\req{Sigma_e_RDE}. 
		$\Sigma_\rmi =0$ means full certainty in the initial condition. However, the case of no prior 
		information on $x(0)$ means $\sfX=0$ or $\Sigma_\rmi$ somehow infinite. This case cannot be 
		captured by the covariance form, but is easily incorporated in the information/certainty form 
		  since $S(0)=\sfX=0$ is allowed. In this case, it is observability 
		that ensures $S(t)>0$ for $t>0$. 
	\end{remarc}

%
%
%
%
%
%
%
%
%
%
%
%
%
	
	\section{Discussion} 							\label{disc.sec}
	
	It is often stated that controllability and observability are dual notions, or that state-feedback 
	and estimation involve some kind of duality. 
	These various 
	 notions of duality have been mentioned for decades in papers and textbooks. 
	Recent work~\cite{kim2025arrow} has revisited these arguments from the point of view 
	that duality in the mathematical sense is the relation between a problem defined on vectors, 
	and what the same problem means for {\em functionals} on those same vectors. For dynamical 
	systems, duality typically involves a time reversal, but it's not true that just a simple time reversal 
	yields a dual problem. For example, the two problems $\LQRi$ and $\LQRf$ in Section~\ref{LQR.sec} 
	are equivalent modulo a time reversal. However, these problems are not duals in the sense of 
	duality between vectors and functionals. 
	
	It is not compelling to think of the LQR problem 
	and the Kalman filter as dual problems either since their respective structures are quite different. 
	The LQR state-feedback controller is memoryless, while 
	the Kalman filter has dynamics, namely the observer.
	As seen in Section~\ref{smooth.sec}, the least-squares {\em smoothing} 
	problem (rather than the {\em filtering} problem) with prescribed final state estimate 
	 is indeed equivalent to LQ-tracking by a time reversal. 
	This brings up the question of what is the time-reversal of the LSSE problem? 
	This would be an LQ-tracking problem where the initial state is unspecified. 
	Since this 
	 problem has no initial or final  state constraints, it turns out its solution parallels that of  
	 the Kalman filter with the output signal $y$ replaced by the reference signal $r$. 
	
	
	There is one conclusion that can be drawn from the similarity of 
	the LQ-tracking, filtering, and smoothing  problems on the one hand, and their contrast with LQR 
	on the other. The former  have dynamics, while the LQR controller
	is memoryless.  
	The dynamics~\req{po_DRE_LQT} in LQ-tracking, 
		and those~\req{po_DRE_K},~\req{P1_DRE_K_smooth}  in filtering and smoothing
		 all arise from the {\em linear} 
	part of the affine-quadratic cost, while the LQR problem does not have such a term in the cost. 
	
	The emergence of dynamics requires further commentary. 
	It is useful to recall  from Dynamic Programming that the optimal control for any 
	objective of the form 
	\be
		J = \smint{0}{T} \phi\big( x(t) , u(t) , t \big) \, dt  			\label{phi_xu.eq}
	\ee
	is a memoryless state feedback $u(t) = \cK\big( x(t),t \big) $ for some mapping $\cK(.,t)$ from the state space
	to the set where the control takes its values. 
	The objectives in both tracking and estimation are of the form 
	\be
		J = \smint{0}{T} \theta\big( x(t) , u(t) , z(t),  t \big) \, dt 
			 = \smint{0}{T} \phi\big( x(t) , u(t) ,  t \big) \, dt  , 	
		\hstm 
		\phi(x,u,t):= \theta(x,u,z(t),t), 		\label{phi_xuz.eq}
	\ee
	where the minimization is over the signals $x,u$, and $z$ is a fixed known signal ($r$ in the case 
	of tracking, and $y$ in the case of filtering/smoothing). The optimal control $u$ is still of the
	form of memoryless state feedback, e.g. in the tracking problem it is~\req{LQT_sol} 
	\be
		u(t) \speq -R^{-1}B^* \big( P(t)\, x(t) + p_1(t) \big) \, =: \, \cK\big( x(t) , t \big) .	\label{u_sum.eq}
	\ee
	Although $\cK(.,t)$ is a memoryless mapping $x(t) \mapsto u(t)$, 
	it does depend on $p_1$, which is formed 
	from the signal $r$ by a dynamical system (anticausal in this case). Thus it is the mapping 
	$r \mapsto \cK$ that has (future) memory. 
	
	In the smoothing problem,
	the optimal  ``control input'' 
	is actually the optimal process disturbance $\wh$, which is obtained from $\xh_\rms$ by 
	Equation~\req{wh_diffeq} 
	\[
			\wh(t)  \speq \sfW^{-1} S(t) \, \xh_\rms(t)  + \sfW^{-1} \, s_1(t)  
			\, =: \, \cK\big( \xh_\rms(t) , t \big) 
	\]
	The mapping $\xh_\rms(t) \mapsto \wh(t)$ above  is indeed  memoryless, but again it is the mapping 
	$s_1\mapsto \cK$ that has the (past) memory of the signal $y$. Note that $\xh_\rms$ is the smoother's
	 state 
	that evolves backwards in time, while $x$ in the tracking control~\req{u_sum}
	 is the system's state that evolves 
	forward in time.

%
	
%
	
	One advantage of the  deterministic formulation of state estimation is that it generalizes to 
	infinite-dimensional systems more easily than the stochastic version. The stochastic version 
	requires attention to the many technicalities of infinite-dimensional random vectors and random 
	fields~\cite{Balakrishnan1981}. For spatially-distributed systems, the matrices $\sfW,\sfV,\sfX$ in 
	the deterministic
	 problem formulation~\req{l_uncert} become spatial operators that describe spatial relations between 
	the uncertainties $w,v,x(0)$ respectively. Spatially-distributed estimation problems can then be 
	tackled in a similar manner to this paper using the machinery of infinite-dimensional 
	optimal control, which is 
	well established~\cite{banks1983state,bensoussan2007representation,curtain2020introduction} 
	in comparison to the stochastic case. In particular, issues such as
	 Ito versus Stratonovic interpretations of 
	stochastic Partial Differential Equations (PDEs) would be completely avoided. 
	
	Finally we point out that the time-reversal argument between $\LQRi$ and $\LQRf$  in Section~\ref{LQR.sec} 
	applies just as well to nonlinear dynamics and nonquadratic objectives. 
	The time-reversed dynamics would be 
	\[
		\begin{aligned} 
			\xd &= f(x,u), \\ 
			x(\sT) & =\xfin, 
		\end{aligned} 
		\hstm \Leftrightarrow \hstm 
		\begin{aligned} 
			-\dot{\xt} &= f(\xt,\ut), \\ 
			\xt(0) & =\xfin, 
		\end{aligned}  
		\hstm  
		\begin{aligned} 
			\xt(t) :=&\, x(\sT-t), \\ 
			\ut(t) :=&\, u(\sT-t).
		\end{aligned} 
	\]
	The corresponding 
	cost-to-go and cost-to-arrive value functions would simply be the time reversals of each other.

\bibliographystyle{IEEEtran}
\bibliography{Det_Kalman}
\end{document}